\documentclass[fleqn,usenatbib]{mnras}

\usepackage{newtxtext,newtxmath}
\usepackage[T1]{fontenc}

\DeclareRobustCommand{\VAN}[3]{#2}
\let\VANthebibliography\thebibliography 
\def\thebibliography{\DeclareRobustCommand{\VAN}[3]{##3}\VANthebibliography}

\usepackage{graphicx}    % Including figure files
\usepackage{amsmath}     % Advanced maths commands
\usepackage{subcaption}

\usepackage{graphicx}    % Including figure files
\usepackage{amsmath}     % Advanced maths commands
\usepackage{xspace}
\usepackage[utf8]{inputenc}
\DeclareUnicodeCharacter{0300}{\`{}}

\newcommand{\kmsmpc}{\kms\;{\rm Mpc}^{-1}}

\newcommand{\HI}{\ion{H}{i}\xspace}
\newcommand{\Htwo}{\ensuremath{\text{H}_2}\xspace}

\newcommand{\hkpc}{h^{-1}{\rm kpc}}
\newcommand{\hmpc}{h^{-1}{\rm Mpc}}

\newcommand{\kms}{\;{\rm km}\,{\rm s}^{-1}}

\newcommand{\gad}{{\sc Gadget-3}}

\newcommand{\simba}{\mbox{{\sc Simba}}\xspace}
\newcommand{\gizmo}{\mbox{{\sc Gizmo}}\xspace}
\newcommand{\mufasa}{\mbox{{\sc Mufasa}}\xspace}

\title[Machine Learning on Simulations for Intensity Mapping]{Populating Galaxies Into Halos Via Machine Learning on the \simba Simulation}

\author[P. K. Das et al.]{
Pratyush Kumar Das,$^{1,2,3}$\thanks{E-mail: pkd45.phy@gmail.com}
Romeel Davé,$^{2,6}$
Weiguang Cui$^{2,4,5}$
\\
$^{1}$School of Mathematics and Physics, University of Queensland, Brisbane, QLD 4072, Australia\\
$^{2}$Institute for Astronomy, Royal Observatory, Edinburgh EH9 3HJ, UK\\
$^{3}$School of Physical Sciences, National Institute of Science Education and Research, HBNI, Jatni 752050, Odisha,
India\\
$^{4}$Departamento de F\'{i}sica Te\'{o}rica, Universidad Aut\'{o}noma de Madrid, M\'{o}dulo 15, E-28049 Madrid, Spain  \\
$^{5}$Centro de Investigaci\'{o}n Avanzada en F\'isica Fundamental (CIAFF), Facultad de Ciencias, Universidad Aut\'{o}noma de Madrid, 28049 Madrid, Spain \\
$^{6}$Department of Physics and Astronomy, University of the Western Cape, Bellville, 7535 Cape Town, South Africa
}

\date{Accepted XXX. Received YYY; in original form ZZZ}

\pubyear{2015}

\begin{document}
\label{firstpage}
\pagerange{\pageref{firstpage}--\pageref{lastpage}}
\maketitle

\begin{abstract}
We present a machine learning (ML) based framework, \textsc{Machine Inferred Galaxy} (MIG), designed to populate dark matter halos with galaxies in N-body simulations. MIG predicts galaxy stellar mass ($M_*$), star formation rate (SFR), atomic and molecular gas masses (\HI\ mass ($M_{\rm HI}$), and H$_2$ mass ($M_{\rm H2}$)), and metallicity, and can be readily extended to other galaxy properties and simulations. The framework first separates halos into central and satellite systems, then uses ML classifiers to distinguish star-forming (SF) from quenched (Q) galaxies, followed by separate regressors trained on the SF subgroups for both centrals and satellites. MIG is trained on the $(100\,h^{-1}\mathrm{Mpc})^3$ \simba\ galaxy formation simulation at $z=0$ and achieves high accuracy for key baryonic properties, including a regression score close to 0.9 for $M_{\rm HI}$ predictions of central galaxies. We further demonstrate its robustness at $z=1$ and $z=2$. Training on fractional quantities (e.g. $M_{\rm HI}/M_*$) and then rescaling by the predicted $M_*$ yields improved performance over direct predictions across all properties and redshifts. MIG also reproduces galaxy mass distribution functions with higher fidelity, an essential step for accurately predicting integrated quantities such as \HI\ intensity maps. These results establish MIG as an efficient and physically consistent tool for generating mock galaxy catalogues and baryonic tracers in large cosmological volumes for various surveys.
\end{abstract}


%%%%%%%%%%%%%%%%%%%%%%%%%%%%%%%%%%%%%%%%%%%%%%%%%%

%%%%%%%%%%%%%%%%% BODY OF PAPER %%%%%%%%%%%%%%%%%%

\section{Introduction}
\label{sec:Intro}

The connection between galaxies and their host dark matter halos has long been one of the most important areas of research in cosmology and galaxy evolution. Halos are essentially the building blocks of large-scale structures, where baryonic matter condenses to form individual galaxies, groups, and clusters. Large cosmological N-body simulations can self-consistently model the evolution of halos and give us detailed insight into their abundance, clustering, and substructure in a $\Lambda$CDM universe.

In contrast to halos, simulations face greater challenges in modeling the baryonic content of galaxies~\citep{somerville2015physical, naab2017theoretical}. Cosmological hydrodynamic simulations such as \simba~\citep{dave2019simba}, Illustris~\citep{vogelsberger2014introducing}, EAGLE~\citep{schaye2015eagle}, and \mufasa~\citep{10.1093/mnras/stw1862} use state-of-the-art models to reproduce numerous properties observed in the galaxy population. Such simulations enable direct investigations of the intrinsic connection between dark matter and baryons. However, generating such simulations with sufficiently large volumes and high enough resolution to properly model galaxies is greatly limited by computational cost. Modern simulations can reach scales of up to hundreds of Mpc on a side, but this is still well short of the $\gtrsim$Gpc$^3$ boxes available in N-body simulations, which are necessary to explore precision cosmological constraints on dark energy and modified gravity. Approximate methods such as semi-analytic models (SAMs) can populate halos with galaxies~\citep[e.g.][]{benson2012galacticus}, but are based on the assumption that baryons follow the dynamics of dark matter \citep[see][for the effect of baryons, for example]{Cui2014}.

A modern data-science-based approach to populating galaxies into halos takes advantage of emerging ML techniques. These ML approaches learn the relations between dark matter halo properties and galaxy properties based on hydrodynamic simulations and then generate galaxies from the N-body simulations in a supervised manner~\citep{horowitz2024differentiable, de2023machine, chittenden2023modelling, hausen2023revealing, 10.1093/mnras/stad015}. Once effectively trained, this method can be scaled up to much larger volumes in N-body simulations with significantly reduced computational cost.

\cite{kamdar2016machine} applied this approach to galaxies in the Illustris simulation and found that ML could approximately mimic galaxies evolved in an N-body + hydrodynamical simulation. Furthermore, ML can predict the population of galaxies in a few minutes, in contrast to the millions of CPU hours required for the simulations, highlighting its potential for statistical galaxy formation studies. \cite{10.1093/mnras/sty1169} developed a similar framework based on the \mufasa\ simulation; however, it was only applied to pre-selected central SF galaxies, since including satellites or Q galaxies resulted in significantly poorer predictions. \cite{10.1093/mnras/stx3293} and \cite{10.1093/mnras/staa234} applied an ML framework to predict \HI\ data from broadband optical properties; while it did not connect galaxies to halos, it introduced the idea of using an ML pre-classifier to distinguish SF from Q galaxies. \cite{lovell2022machine} addressed the limitation that the training set from small-volume hydrodynamic simulations does not include large galaxy clusters by employing a combined training set of the EAGLE~\citep{schaye2015eagle} simulation plus the C-EAGLE~\citep{10.1093/mnras/stx1647} cluster zoom simulations. \cite{10.1093/mnras/stab1120} employed a hybrid approach where the SFR was predicted using the analytic equilibrium model~\citep{10.1111/j.1365-2966.2011.20148.x, mitra2015equilibrium}, then used as an input for ML to train on gas properties, improving accuracy. In many of these works, the ML framework does an excellent job of predicting quantities that grow steadily (like $M_*$ or metallicity) but struggles with accuracy for features that fluctuate over time (SFR, $M_{\rm HI}$, and $M_{\rm H2}$). Many upcoming large-scale structure and intensity mapping surveys for cosmology employ tracers closely related to the latter quantities. Thus, while ML has great potential to populate halos with baryonic properties, there remains substantial room for improvement.

In this paper, we build on these previous works to develop an ML framework, that learns the connection between key galaxy baryonic properties and their host dark matter halo properties, and which can be used to populate an entire N-body simulation. Our goal is to develop an accurate framework for predicting a wide range of galaxy properties, especially SFR and gas content. The previous works discussed above focused primarily on predicting central galaxy properties, which are most closely connected to halo assembly. However, populating a full set of galaxies in an N-body simulation also requires handling both centrals and satellites (i.e. subhalos in an N-body simulation), with satellites following different relations since they are subject to additional processes such as tidal and ram-pressure stripping. Moreover, previous frameworks have generally struggled to automate the dichotomy between SF and Q systems, particularly around $L^\star$. Thus, bringing these ML techniques to maturity requires an end-to-end pipeline to optimally predict a wide range of properties for the full galaxy population from dark matter information alone.

\begin{figure}
    \centering
    \includegraphics[width=\columnwidth]{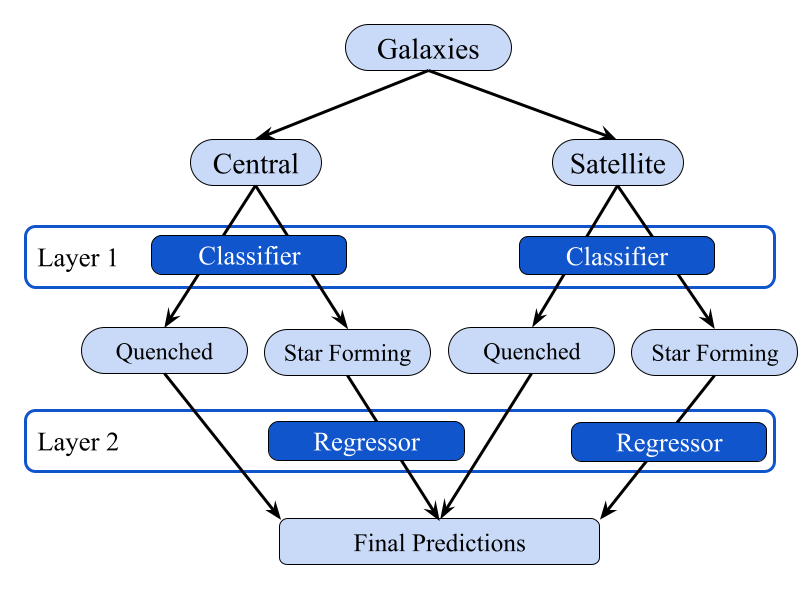}
    \caption{In this flowchart, we show the workflow of MIG. Starting from the top, galaxies are separated into central and satellite subgroups. Following this, we introduce ML classifiers in layer 1 to classify each subgroup into SF and Q galaxies, based on the relevant galaxy feature. Subsequently, distinct regressors are employed in layer 2 to independently train on the SF galaxies, encompassing both centrals and satellites. The Q galaxies are assigned a constant value below the SF–Q boundary and merged with the SF predictions to yield the final results.}
    \label{fig:flow}
\end{figure}

We introduce Machine Inferred Galaxy (MIG), a framework to predict various galaxy properties from the halos and subhalos in an N-body simulation. We train and test our ML framework using the \simba simulation. MIG first separates galaxies into centrals and satellites using the {\sc Caesar}\footnote{\url{ caesar.readthedocs.io}} catalog, as shown in \autoref{fig:flow}. Next, MIG has two main layers: Layer 1 (darker blue) and Layer 2 (lighter blue). At Layer 1, we introduce ML classifiers to classify both central and satellite galaxies into Q and SF subgroups. After the classification, in Layer 2, we introduce ML regressors that independently train on the SF subgroups of central and satellite galaxies. At this stage, the mean value for each feature from the Q galaxies is assigned to all final predictions for Q galaxies. The main motivation for having a classification for SF and Q groups was to remove the extreme outliers (Q galaxies) from the training dataset of the SF regressor and increase its accuracy. Ultimately, we merge the predictions from the regressors with the constant values assigned to Q, resulting in predictions for all galaxies in the simulation. We explore a large variety of ML algorithms, assembling an optimal combination of algorithms for both layers. We then apply this ML framework to an N-body version of the \simba simulation, specifically toward the use case of making predictions for the \HI\ 21cm intensity mapping.

\begin{figure}
    \centering
    \includegraphics[width=\columnwidth]{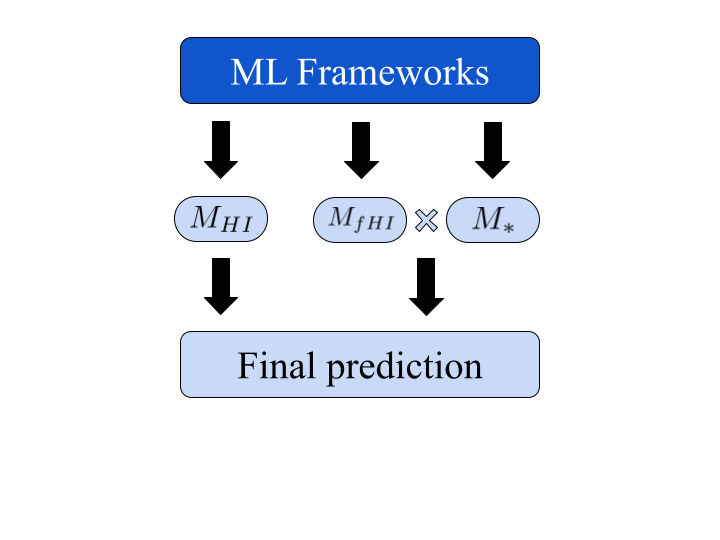}
    \vskip-0.5in
    \caption{An illustration of the two approaches we follow to predict the target galaxy features. Here we show the example of $M_{HI}$ prediction. The left half of the figure shows the traditional approach, where ML frameworks are trained to predict $M_{HI}$ directly. The right half shows the fraction-based approach, where we train two ML frameworks for $M_{fHI}$ ($=M_{HI}/M_{*}$) and $M_{*}$ respectively. Ultimately, we multiply them to get back $M_{HI}$ or the main target feature.}
    \label{fig:twoapproach}
\end{figure}

Another unique aspect of MIG is that we employ feature modifications to improve accuracy. Various previous works~\citep{lovell2022machine,de2022mimicking,10.1093/mnras/sty1169} have shown that ML algorithms can predict $M_*$ accurately. To incorporate this advantage, we divided the other relevant galaxy features by $M_*$ and trained the ML models to predict these new features. Thus MIG is trained to predict the \HI\ fraction ($f_{\rm HI} = M_{\rm HI}/M_{*}$), the H$_2$ fraction ($f_{\rm H2} = M_{\rm H2}/M_{*}$), and the specific star formation rate ($\mathrm{sSFR} = \mathrm{SFR}/M_{*}$). We then multiply them by the $M_*$ predictions from MIG to recover the desired galaxy features ($M_{\rm HI}$, $M_{\rm H2}$, and SFR), as shown in \autoref{fig:twoapproach} for $M_{\rm HI}$ predictions. Although this process employs two distinct ML frameworks; the first to predict the fraction and the second to predict $M_*$, we show in \autoref{sec:results} that it usually improves the accuracy of the predictions. Overall, MIG predicts eight distinct features, each with different pipelines: $M_{*}$, $M_{\rm HI}$, $f_{\rm HI}$, $M_{\rm H2}$, $f_{\rm H2}$, SFR, sSFR, and metallicity ($Z$).

Ultimately, a key goal of populating galaxies into dark matter halos is to accurately determine the baryonic mass distributions across the constituent galaxies. This is critical for most science cases, including creating mock surveys with selection functions or recovering the overall mass within large-scale pixels for intensity mapping. The way to quantify this aspect is via the mass function. MIG also provides a post-processing stage to compute and compare these functions for both our approaches, to finalize the strategy for an accurate prediction of intensity maps.

In \autoref{sec:simba} of this paper, we briefly review the \simba simulation and describe the input and output parameters of our ML frameworks. Our approach to using the participating ML algorithms and the workflow of MIG is explained in \autoref{sec:mls}. \autoref{sec:results} shows predictions of MIG and compares them with the \simba values using performance metrics. We also show the applicability of MIG at $z=1,2$ through SFR, $M_{\rm HI}$, and $M_{\rm H2}$ predictions in \autoref{sec:regz}. For each redshift, the mass functions for each feature are presented in \autoref{sec:massfunc}.

%%%%%%%%%%%%%%%%%%%%%%%%%%%%%%%%%%%%%%%%%%%%%%%%%%%%%%%%%%%%%%%%%%%%%%%%%%%%%%%%%%%%%%%%%%%%%%%%%%%%%%%%
%%%%%%%%%%%%%%%%%%%%%%%%%%%%%%%%%%%%%%%%%%%%%%%%%%%%%%%%%%%%%%%%%%%%%%%%%%%%%%%%%%%%%%%%%%%%%%%%%%%%%%%%

\section{THE SIMBA SIMULATION}
\label{sec:simba}

The \simba simulation is a cosmological hydrodynamic simulation with state-of-the-art galaxy formation modules. It is built on the \gizmo\ code~\citep{hopkins2015new}, which uses meshless finite mass hydrodynamics, with the gravity solver mostly taken from \gad~\citep{springel2005cosmological}. The input physics includes gravity, radiative cooling (including metal lines), and on-the-fly self-shielding from a spatially uniform photo-ionizing background, which directly predicts the \HI\ fraction in each gas particle. Additional processes include star formation based on an on-the-fly subgrid H$_2$ model, stellar feedback using decoupled kinetic winds, black hole growth using both torque-limited and Bondi accretion, black hole feedback in three different modes, and on-the-fly dust production and destruction. Full details of these modules are available in \cite{dave2019simba}. This comprehensive set of input physics generates galaxies and intergalactic gas properties that align well with a wide range of observations, the most relevant for this work being the gas content of galaxies as a function of mass \citep{10.1093/mnras/staa1894} and the Q fraction of galaxies as a function of mass \citep{dave2019simba}. Thus, \simba\ represents a plausible (though not unique) model for the growth and evolution of galaxies from early epochs to the present day.

The main \simba\ simulation from the \simba suite, and the one used in this work, is run in a $(100\hmpc)^3$ comoving volume, with an adaptive Plummer-equivalent gravitational softening length of $0.5\hkpc$. The gas particle mass resolution is $1.82\times 10^7M_\odot$, and the dark matter particles have a mass of $9.8\times 10^7M_\odot$. The assumed cosmology is consistent with Planck+2015, with $\Omega_m=0.3$, $\Omega_\Lambda=0.7$, $\Omega_b=0.048$, $H_0=68\kmsmpc$, $\sigma_8=0.82$, and $n_s=0.97$.

We identify halos on the fly with the Friends-of-Friends (FoF) method, incorporated into \gad. Within each halo, galaxies are identified as objects containing stars and dense gas, using a 6D FoF finder incorporated in the {\sc Caesar} simulation analysis package. Cold gas (\HI\ and \Htwo) is associated with galaxies within each halo as belonging to the galaxy to which it is most bound, allowing \HI, in particular, to extend beyond the interstellar medium. In this way, {\sc Caesar} galaxies contain essentially all ($>98\%$) of the \HI\ in the simulated universe at $z \leq 2$. In this work, we use snapshots 151, 105, and 78 at $z \approx 0,1,2$, respectively. Note that \simba data, including particle snapshots and associated {\sc Caesar} catalogs containing hundreds of pre-computed properties for each galaxy and halo, is publicly available at \url{simba.roe.ac.uk}.

%%%%%%%%%%%%%%%%%%%%%%%%%%%%%%%%%%%%%%%%%%%%%%%%%%%%%%%%%%%%%%%%%%%%%%%%%%%%%%%%%%%%%%%%%%%%%%%%%%%%%%%%

\subsection{Halo and Galaxy Properties}
\label{sec:properties}
We use the halo and galaxy properties of the \simba\ 100 Mpc/h box taken from the {\sc Caesar} catalogs at specific redshifts. Selecting suitable features before training a ML model is crucial for ensuring model accuracy, reducing overfitting, and improving interpretability \citep{li2017feature}. The feature selection step in MIG removes redundant or unimportant features from all available dark-matter-only halo properties in \simba\ and retains only the most predictive ones. We first train separate Random Forest Classifier/Regressor models (please refer to \autoref{sec:mls} for details) for each galaxy property. The Random Forest model ranks all the input features based on their importance in predicting the target variable. These rankings are stored and used in the next crucial step of our feature selection process.

Many astrophysical quantities exhibit strong correlations with each other. For example, halo mass and velocity dispersion are closely linked \citep{10.1093/mnras/sty590}. Including highly correlated features can introduce multicollinearity, leading to redundant information that may negatively impact the model's interpretability and performance. To mitigate this, we compute the correlation matrix using the publicly available Pandas library and identify features with a correlation coefficient exceeding 0.9. Among the correlated feature pairs, we remove the lower-ranked feature based on the Random Forest rankings, ensuring that only the most relevant and independent features are retained.

We consider the following halo properties from \simba:

\begin{itemize}
    \item \textbf{Radii:} We employ six radii: one that encloses half (50\%) of the dark matter halo mass ($r_{1/2}$), two that contain 20\% ($r_{20}$) and 80\% ($r_{80}$) of the entire dark matter halo mass, and three radii related to the critical density, namely where density reaches 200, 500, and 2500 times the critical density ($R_{200}$, $R_{500}$, $R_{2500}$). We use $r_{1/2}$ and $R_{500}$ for both central and satellite galaxy predictions; through experimentation, we found it is best to include $r_{20}$ and $R_{200}$ only for central galaxies, while only using $r_{80}$ and $R_{2500}$ for satellite galaxies.
    \item \textbf{Environmental mass density:} We use three environmental mass-density estimates of {\sc Caesar} galaxies at various scales, within spherical tophat apertures of 300 kpc, 1000 kpc, and 3000 kpc (comoving).
    \item \textbf{Halo spin:} We use the dimensionless spin ($\lambda$)\citep{bullock2001universal}, defined as:
    \begin{equation}
        \lambda=\frac{J_{\mathrm{vir}}}{\sqrt{2} M_{\mathrm{vir}} R_{\mathrm{vir}} V_{\mathrm{vir}}},
    \end{equation} 
    where $J_{\rm vir}$ is the angular momentum within radius $R_{\rm vir}$, $M_{\rm vir}$ is the virial mass, and $V_{\rm vir}=\sqrt{2 G M_{\rm vir} / R_{\rm vir}}$ is the halo circular velocity.
    \item \textbf{Velocity dispersion:} The mass-weighted velocity dispersion for dark matter particles, computed around the center-of-mass velocity.
    \item \textbf{Angular momentum:} The magnitude of the specific angular momentum vector (angular momentum/halo dark matter mass) for the dark matter component of the halo.
    \item \textbf{Formation redshift:} The redshifts at which the halo accretes 25\%, 50\%, and 75\% of its final mass. For satellite galaxies, we do not include halo formation redshift as an input feature. This is based on our analysis that halo formation redshift does not directly correlate with the growth patterns of satellite galaxies.
    \item \textbf{Subhalo mass:} The sum of the mass of all dark matter particles within a 30 kpc (comoving) spherical aperture. This feature can help in understanding the gravitational influence and potential growth trajectory of the subhalo.
    \item \textbf{Subhalo velocity:} The magnitude of the velocity difference between the subhalo and its host halo. This can provide insights into the dynamical state of the subhalo relative to the center of the halo.
    \item \textbf{Subhalo distance:} The distance of the subhalo from the center of its host halo. This metric can be crucial for studying environmental effects on the subhalo.
\end{itemize}

We present the feature-importance results in \autoref{sec:results}.

Our ML frameworks are trained to predict the following galaxy properties from \simba:
\begin{itemize}
    \item \textbf{Stellar mass ($M_{*}$):} The total mass of all star particles in a galaxy.
    \item \textbf{\HI mass ($M_{\rm HI}$):} Particle \HI fractions are computed on the fly assuming a self-shielding model based on \citet{rahmati2013evolution}. A galaxy's \HI mass is the sum of all \HI in particles that are most bound to it within its halo.
    \item \textbf{\Htwo mass ($M_{\rm H2}$):} Particle \Htwo fractions are computed on the fly using a subgrid model based on \citet{krumholz2011comparison}. The \Htwo mass of a galaxy is computed analogously to the \HI mass.
    \item \textbf{Star formation rate (SFR):} Particle SFRs are computed from the \Htwo density divided by the local dynamical time, multiplied by an efficiency of 0.02. The total SFR is the sum of individual particle SFRs within a galaxy.
    \item \textbf{Metallicities} ($Z$)\textbf{:} We use two types of metallicities: SFR-weighted and $M_{*}$-weighted. Our goal is to predict the SFR-weighted metallicity, but for galaxies without SFR values (generally old galaxies), we predict the $M_{*}$-weighted metallicity, as every galaxy has stellar mass. In general, the difference between SFR-weighted and mass-weighted $Z$ is small.
\end{itemize}

%%%%%%%%%%%%%%%%%%%%%%%%%%%%%%%%%%%%%%%%%%%%%%%%%%%%%%%%%%%%%%%%%%%%%%%%%%%%%%%%%%%%%%%%%%%%%%%%%%%%%%%%

\section{MACHINE LEARNING SETUP}
\label{sec:mls}
Galaxies are categorized into diverse populations, each characterized by distinct attributes. Among these classifications, galaxies are differentiated into central and satellite categories, reflecting their positions in cosmic structures. Furthermore, galaxies are distinguished based on their stellar activity, identified as SF or Q. This bifurcation enables the creation of four specific subgroups: Central-SF, Central-Q, Satellite-SF, and Satellite-Q. As highlighted in \autoref{sec:Intro}, existing ML frameworks have predominantly focused on analyzing the Central-SF subgroup or, if fitting anything else, have struggled to distinguish between these groups effectively.

In the {\sc Caesar} catalog, the galaxy with the highest stellar mass ($M_*$) within a halo is designated as the central galaxy, while all remaining galaxies are classified as satellites. This delineation allows for a straightforward categorization into central and satellite sub-populations. To differentiate between SF and Q galaxies, we follow the methodology outlined in \cite{10.1093/mnras/staa234}, employing an ML classifier for this purpose. However, unlike them, we train our classifiers on the sSFR boundaries taken from \cite{dave2019simba} as:
\begin{equation}
    \mathrm{sSFR}=10^{-1.8+0.3 z} \mathrm{Gyr}^{-1},
    \label{eq:sSFR}
\end{equation}
where $z$ is the redshift. Subsequently, we train the ML classifier on this boundary based on the halo's attributes at redshifts 0, 1, and 2. This approach enables us to automate the identification of central/satellite and SF/Q galaxies from a collection of halos with recognized subhalos. Consequently, the classification of galaxies into four distinct subgroups is seamlessly integrated into our pipeline.

For the Q galaxies, we assume they have negligible SFR, $M_{\rm HI}$, and $M_{\rm H2}$. This assumption generally holds in \simba; the SF galaxy population retains 80.3\%, 85.7\%, and 83.7\% of the overall SFR, $M_{\rm HI}$, and $M_{\rm H2}$, respectively, at $z=0$. For higher redshifts, for $M_{\rm HI}$, the SF population share at $z=1$ is 94.8\%, and at $z=2$ is 98.0\%. Thus, in terms of large-scale observables such as intensity mapping, the Q population adds a minor or small contribution that we ignore. We leave it for future work to correct for the gas and SFR contributions from Q galaxies. Our pipeline automatically outputs the values of the Q galaxy features as a constant value, lower than the SF/Q boundary.

For $M_{*}$, our approach deviates slightly from the previously described framework. While we continue to use the {\sc Caesar} catalog for segregating galaxies into central and satellite categories, the application of ML classifiers to distinguish between SF and Q galaxies is not required in this context. This arises because the demarcation between Q and SF populations based on $M_{*}$ is not as pronounced as for other features. Therefore, for $M_{*}$, we directly implement a regressor model for central and satellite galaxies, focusing on predicting $M_{*}$ without the need to pre-classify galaxies as SF or Q.

For the SF galaxies, we train ML regressors to learn the relation between halo and galaxy features separately for centrals and satellites. Training, testing, and optimization of these regressors are performed for each subgroup and are ultimately consolidated into a cohesive model.

In the subsequent sections, we describe the specifics of the ML algorithms employed for both classification and regression tasks, along with the Random Forest algorithm used for feature selection. A key aspect of the ML algorithms is the presence of adjustable parameters known as ``hyperparameters''. Adjusting these parameters allows for the customization of the model to meet our specific objectives. Manual tuning of these hyperparameters designates the algorithms as ``hand-tuned''. This manual tuning process, while offering precise control, can be labor-intensive and complex. To address this, we explore automated ML algorithms as described below. Automated approaches facilitate the comparison of different ML algorithms by autonomously adjusting their hyperparameters, thereby optimizing them based on a predetermined criterion. After completing this comparative analysis, we assess the accuracy metrics of each method. This evaluation enables us to integrate the strengths of each approach, culminating in the derivation of the most effective ML algorithms, complemented by optimally tailored sets of hyperparameters.

\subsection{Random Forest (RF)}
RF is a decision-tree-based ensemble learning algorithm that excels at handling high-dimensional datasets and effectively capturing non-linear relationships between the input features. Although RF is widely used as a classifier and regressor, we leverage its ability to perform feature selection by ranking input features based on their importance. During training, RF evaluates the contribution of each feature by analyzing its role in reducing impurity across all decision trees in the forest. The average impurity reduction attributed to each feature is then used to compute an importance score, which enables the ranking of features from most to least significant.

We use the Random Forest regressor and classifier implementations from the \texttt{Scikit-Learn} Python library.

\subsection{Tree-based Pipeline Optimization Tool (TPOT)}
{\sc TPOT} \citep{OlsonGECCO2016} is a distinguished AutoML framework that has undergone significant advancements in recent years. We have integrated TPOT’s publicly available Python library into our research pipeline. TPOT relies on the robust {\sc Scikit-Learn} library to handle data and provides access to a comprehensive suite of ML algorithms. The latest version of TPOT introduces a graph-based implementation that supports genetic feature selection, flexible search spaces, and multi-objective optimization, improving both efficiency and modularity. TPOT employs genetic programming, a stochastic global search technique, to autonomously explore an expansive space of potential ML pipelines, ultimately identifying the top-performing model for the given dataset. These enhancements ensure greater flexibility, extendability, and maintainability, making TPOT an even more powerful tool for optimizing ML workflows.

\subsection{Classifier}
\label{subsec:clf}
The aim is to determine whether a galaxy is SF or Q based on sSFR. The dataset is initially split into central and satellite subgroups. At $z=0$, the \simba 100 Mpc/h box contains 35,843 central and 19,766 satellite galaxies; at $z=1$, it has 27,198 centrals and 12,100 satellites; and at $z=2$, it has 23,042 centrals and 7,701 satellites. We then divide both subgroups at each redshift into two parts—training and testing samples—in a 4:1 ratio. We begin by fitting the training data with a TPOT classifier and later test on the testing sample using parameters derived from the confusion matrix presented below.

\subsubsection{Confusion Matrix}
A confusion matrix describes the performance of a classifier. In our case of distinguishing between two classes considered as positive (SF) and negative (Q), classification yields four possible outcomes:
\begin{itemize}
    \item \textbf{True Positive (TP):} Correct prediction of the positive class.
    \item \textbf{True Negative (TN):} Correct prediction of the negative class.
    \item \textbf{False Positive (FP):} Incorrect prediction of the positive class.
    \item \textbf{False Negative (FN):} Incorrect prediction of the negative class.
\end{itemize}

A key metric used to assess overall classification performance is the \textbf{F1 Score}, which balances precision and recall, making it particularly effective when class distributions are imbalanced. The F1 score is defined as:
\begin{equation}
    \text{F1 Score} = 2 \times \frac{\text{Precision} \times \text{Recall}}{\text{Precision} + \text{Recall}},
\end{equation}
where
\begin{equation}
    \text{Precision} = \frac{TP}{TP + FP}, \hspace{0.2 in}
    \text{Recall} = \frac{TP}{TP + FN}.
\end{equation}

Given the predominance of SF galaxies (positive class) over Q galaxies (negative class) in the \textsc{simba} dataset, high accuracy in predicting the positive class is anticipated because of their substantial representation in the training data. However, our primary objective is to ensure that both SF and Q galaxies are accurately classified. As such, prioritizing the maximization of the F1 score ensures that the classifier maintains a balance between correctly identifying both classes while minimizing false predictions. Along with these, we also show the fundamental accuracy metric, which is the ratio of correct predictions to the entire sample.

\subsection{Regressor}
\label{subsec:reg}
In the previous subsection, we explained the methodology for identifying the most effective classifiers for each of the seven galaxy features. Here, we develop regressor models for the SF subgroup of these features, as well as for $M_*$. We start by using the results from the TPOT Classifier to sort the entire dataset into SF and Q galaxies. Because of the inherent imperfections of these classifiers, some Q galaxies may be misclassified as SF, and vice versa. We aim to enhance the resilience of the regressor models by including SF galaxies that were misclassified as Q in the training dataset. This augmented SF dataset, identified both manually and through ML classification, is then used to train the regressor models. The performance of the regressor is subsequently assessed on the subset of SF galaxies identified solely through ML classification.

We introduce feature-wise TPOT regressors for assessing the accuracy of regressor models. Initially, we visualize the relationship between predicted and true values by plotting the former against the latter for the test dataset. In a perfectly accurate model, all data points in the scatter plot would align with the $y=x$ line, indicating a one-to-one correspondence between true and predicted values. Deviations from this ideal are represented by points scattering around the $y=x$ line, with $y \neq x$.

To quantitatively measure these deviations, we compute five key parameters, analogous to \cite{10.1093/mnras/sty1169}: mean absolute deviation ($\mu$), mean deviation ($\beta$), regression score ($R^2$), the Pearson correlation coefficient ($\rho$), and the slope ($\kappa$) of a linear fit to the scatter plot. We use the publicly available {\sc linregress} package from the {\sc scipy} library to generate the linear fit. $\kappa > 1$ indicates overprediction at the high end of the scatter plot and underprediction at the low end, while $\kappa < 1$ indicates the opposite. The mathematical formulations for these metrics, considering $x_i$ and $\hat{x}_i$ as the logarithms of the true and predicted values for the $i^{\mathrm{th}}$ data point in a sample of size $N$, are:
\begin{equation}
\mu (x, \hat{x}) = \frac{1}{N} \sum_{i=1}^N \left| x_i - \hat{x}_i \right|
\end{equation}

\begin{equation}
\beta (x, \hat{x}) = \frac{1}{N} \sum_{i=1}^N \left( x_i - \hat{x}_i \right)
\end{equation}

\begin{equation}
R^2 (x, \hat{x}) = 1 - \frac{\sum_{i=1}^N \left(x_i - \hat{x}_i\right)^2}{\sum_{i=1}^N \left(x_i - \langle x \rangle\right)^2}
\end{equation}

\begin{equation}
\rho (x, \hat{x}) = \frac{\sum_{i=1}^N \left(x_i - \langle x \rangle\right)\left(\hat{x}_i - \langle \hat{x} \rangle\right)}{\sqrt{\sum_{i=1}^N \left(x_i - \langle x \rangle\right)^2} \sqrt{\sum_{i=1}^N \left(\hat{x}_i - \langle \hat{x}\rangle\right)^2}}
\end{equation}

where $\langle x \rangle = \sum_i^N x_i / N$ and $\langle \hat{x} \rangle = \sum_i^N \hat{x}_i / N$. These metrics are computed for all TPOT regressors across features and redshifts.

%%%%%%%%%%%%%%%%%%%%%%%%%%%%%%%%%%%%%%%%%%%%%%%%%%%%%%%%%%%%%%%%%%%%%%%%%%%%%%%%%%%%%%%%%%%%%%%%%%%%%%%%%%

\section{MIG Framework and accuracy}
\label{sec:results}

The first half of our ML framework is a TPOT classifier trained on sSFR-derived labels of Q and SF, and the second half is a TPOT regressor trained on the SF group to predict the relevant galaxy features.

In this section, we present the TPOT results for each feature under consideration and provide a comparative analysis between the predictions generated by the MIG frameworks and the actual values from the \simba simulation.

\begin{figure}
  \centering
  \vspace{0.35cm}
  \includegraphics[width=\linewidth]{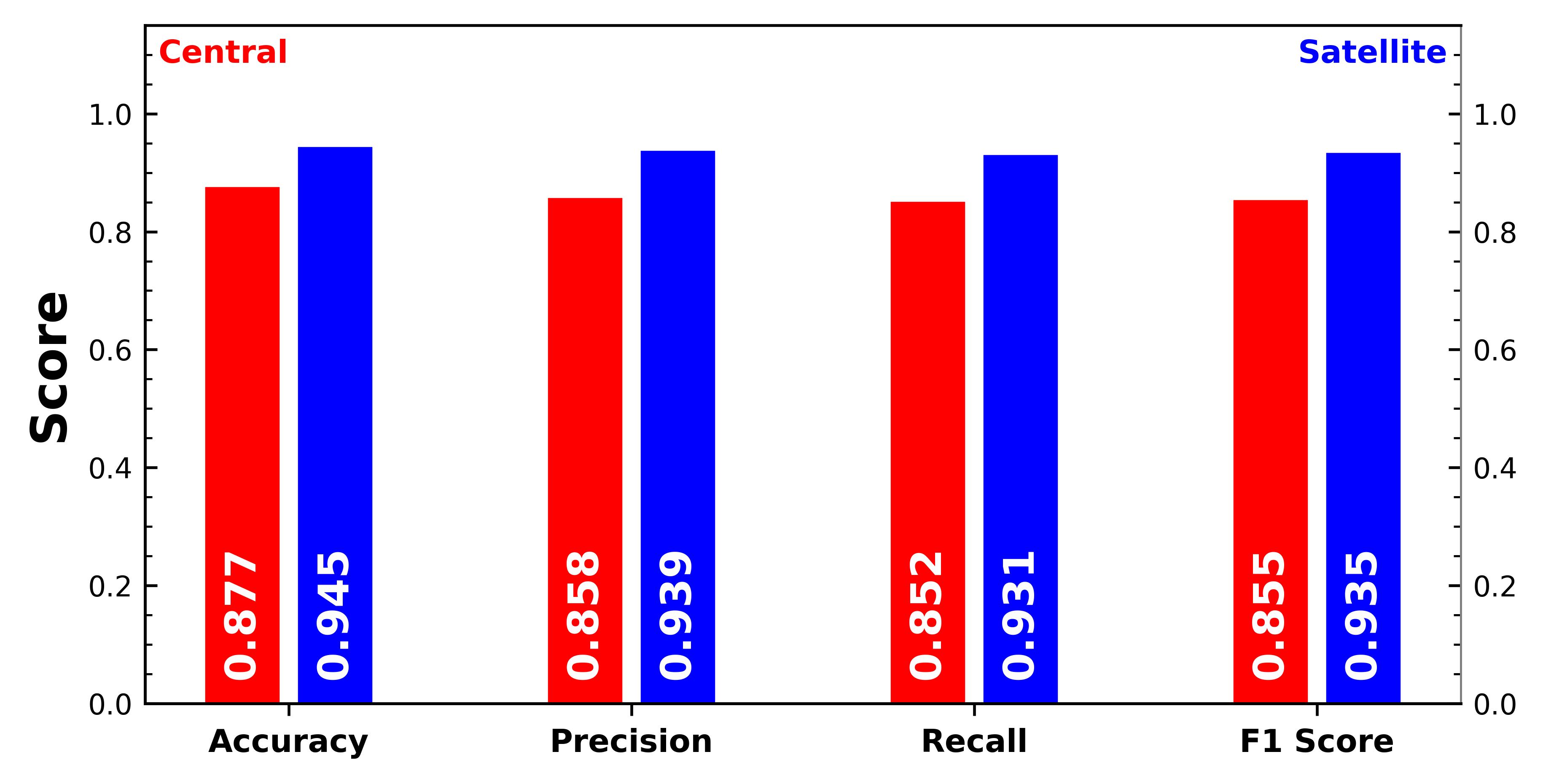}
  \vspace{0.23cm}
  \caption{Performance metrics of the classifier layer of MIG for central (red) and satellite galaxies (blue) at $z=0$. The scores corresponding to each metric are shown inside their respective bars.}
  \label{fig:clf0}
\end{figure}

In Fig.~\ref{fig:clf0}, we present the classifier performance for both central and satellite subgroups. Each metric ranges from 0 to 1, with 1 indicating perfect accuracy. For satellite galaxies, MIG attains an accuracy of 0.945 and an F1 score of 0.935, reflecting very strong predictive performance. For central galaxies, the accuracy drops to 0.877 with an F1 score of 0.855. This reduction is largely from the imbalance inside each subgroup: while about 43\% of satellites are Q, only 14\% of centrals are Q, making the central dataset more biased. Despite this imbalance, the accuracy for centrals remains impressive, with precision and recall values that are nearly balanced. This indicates that the model is handling both the SF and Q populations consistently.

The next part of MIG constructs regressors on the SF subgroup. To illustrate ML accuracy, this section introduces several scatter plots that show the true versus ML-predicted values for various properties from \simba. Notably, these plots will selectively feature galaxies that are classified as SF by both the ML models and the \simba simulation. For completeness, we explain the methodology for constructing such plots in \autoref{sec:app1}.

\begin{figure*}
\begin{subfigure}{.48\textwidth}
  \centering
  \includegraphics[width=1\linewidth]{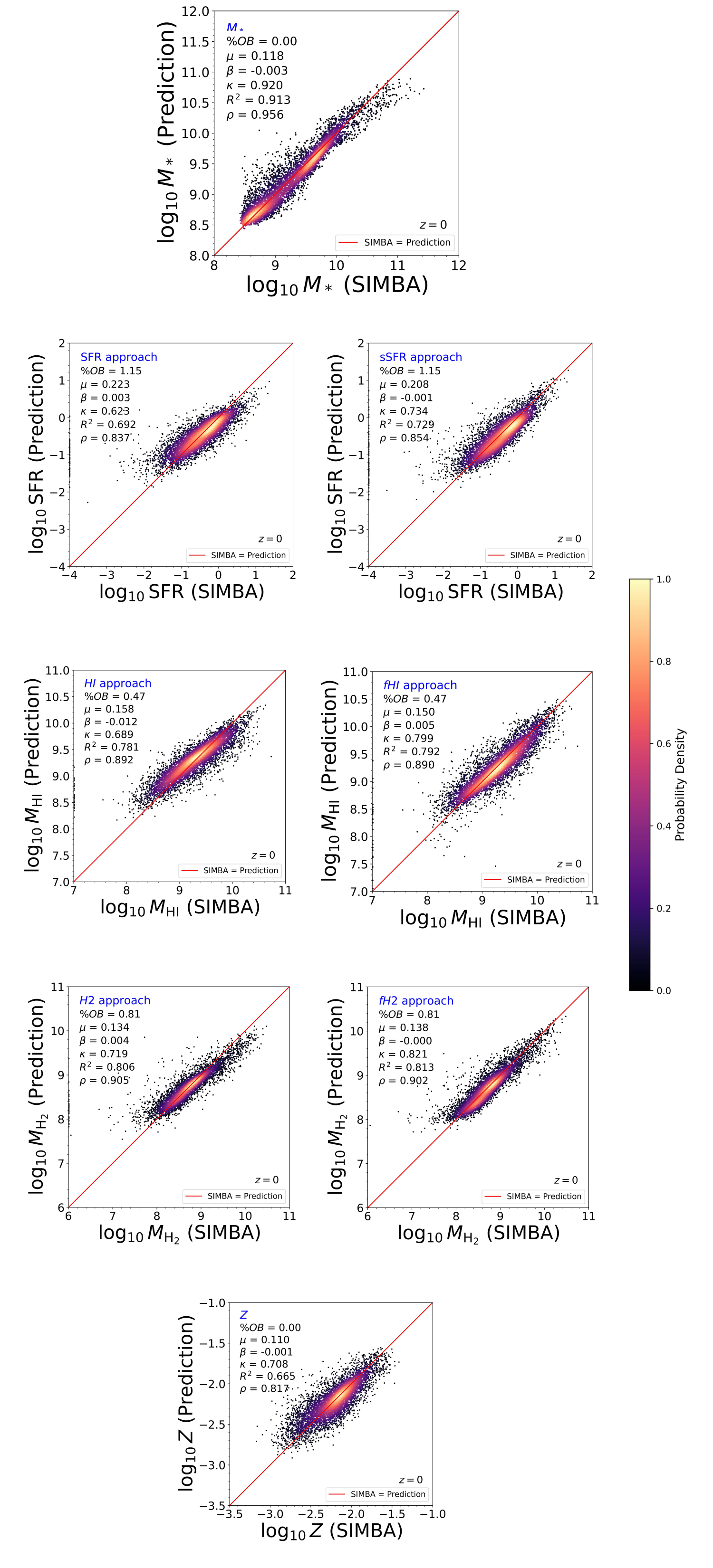}
  \caption{Central Galaxies}
  \label{fig:rmse_scattera}
\end{subfigure}
\begin{subfigure}{.48\textwidth}
  \centering
  \includegraphics[width=1\linewidth]{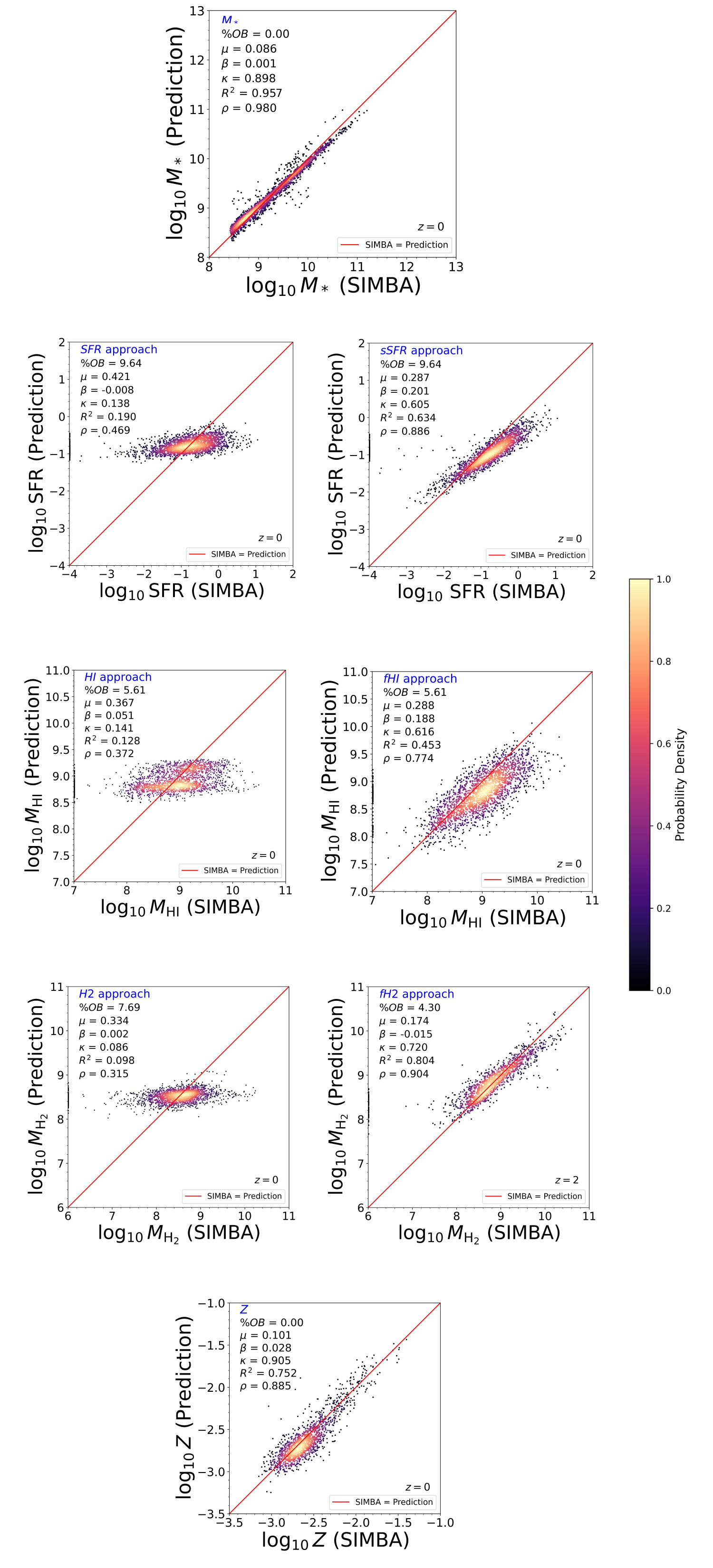}
  \caption{Satellite Galaxies}
  \label{fig:rmse_scatterb}
\end{subfigure}
\caption{Prediction accuracy for the best-performing frameworks for fractions-based and feature-based approaches using the \simba simulation at $z=0$. The scatter plots have the true (\simba) values on the x-axis, while the y-axis shows the predicted values. (a) has ML predictions for central galaxies, and (b) has ML predictions for satellite galaxies.}
\label{fig:rmse_scatter}
\end{figure*}

\autoref{fig:rmse_scatter} shows the accuracy of our ML model predictions (y-axis) compared to the actual values derived from the \simba simulation (x-axis) for each galaxy feature. Each point represents an individual galaxy, offering a detailed view of the model’s predictive accuracy. Ideally, a perfectly predicted value would align with the diagonal red line, indicating a one-to-one correspondence between the predicted and true values. Color coding is used to represent probability densities, which helps distinguish data concentration across the plot. In regions of lower density, individual galaxies are shown in black, clarifying areas with more dispersed data.

Key metrics that quantify the quality of the fit appears in the upper left corner, along with a parameter denoted $\%OB$. This parameter addresses the issue of misclassifications by the classifier, which often predicts values significantly lower than the SF–Q threshold for the respective parameters. Incorporating all data points in the plot tends to dilute focus on high-density regions by broadening the plot boundaries. To counteract this and better highlight high-density areas, we adjust the scatter plot boundaries and designate outliers, typically constituting less than 1\% of the SF subgroup, at the plot’s boundaries. The $\%OB$ parameter quantifies the percentage of such outliers within the total dataset represented in the plot, providing a measure of how these outliers affect the overall data visualization.

In the following subsections, we discuss the fitting accuracy (\autoref{fig:rmse_scatter}) at $z=0$.

\subsection[Stellar mass Stellar mass]{Stellar mass $M_{*}$}
\label{sec:sm}
$M_*$ is generally closely correlated with halo mass in both observations~\citep{more2011accurate} and simulations, with the scatter correlated with halo properties~\citep{cui2021origin}. Thus, it is unsurprising that ML frameworks have generally succeeded in reproducing $M_*$ from dark matter properties~\citep{kamdar2016machine,10.1093/mnras/sty1169}. Our ML framework produces remarkably accurate results, but here we further distinguish between centrals and satellites. As mentioned, $M_{*}$ does not require a classifier to separate SF from Q galaxies.

For both central and satellite galaxies, we have an unbiased distribution across the $y=x$ line ($\beta \approx$ 0 and $\kappa \approx$ 0.9). Although we see marginal underprediction towards the higher mass end of the plot, given it's not a high density region, MIG managed to secure $R^2$ of 0.91 for central and 0.96 for satellite galaxies, as shown in the top panel of \autoref{fig:rmse_scattera}. This demonstrates the efficacy of ML in accurately estimating the stellar masses of central galaxies, supporting findings from previous studies and holding throughout the mass range probed by \simba.

In summary, our analysis confirms that $M_{*}$ can be reliably inferred from dark matter properties using MIG. In future, we plan to extend MIG by incorporating additional cosmological hydrodynamic simulations such as Illustris~\citep{vogelsberger2014introducing} and EAGLE~\citep{schaye2015eagle}, significantly increasing the sample size and testing ground, particularly for satellite galaxies, where our current dataset is limited.

\subsection{Star formation rate SFR}
\label{sec:sfr}
Predicting SFR from dark matter features in simulations using ML has historically been more challenging \citep{10.1093/mnras/sty1169,jo2019machine}. Unlike $M_*$, which is a cumulative quantity and mostly grows (modulo stellar mass loss and satellite stripping), SFR can fluctuate both up and down on timescales significantly shorter than halo dynamical times, owing to local processes such as stochastic gas accretion and internal dynamical evolution. Additionally, satellites can suffer environmental quenching processes that may not correlate directly with dark matter. Yet, predicting SFR is crucial for connecting N-body predictions with, for instance, emission-line surveys. Here we examine the performance of our ML framework in predicting SFR.

Unlike the prediction of stellar mass $M_*$, SFR (along with $M_{\rm HI}$ and $M_{\rm H2}$) requires a preclassifier, depicted in the classifier layer of \autoref{fig:flow}.

With the classifier in place, we next examine the regressor performance, shown in \autoref{fig:rmse_scatter} (a, b), second-row left panels. In this setup, galaxies classified as Q are assigned the mean SFR of the Q population, $10^{-3.73},\mathrm{yr}^{-1}$, while the SF galaxies have their SFRs from the regressor layer predictions of MIG. For both centrals and satellites, the regression results yield $\beta \approx 0$, indicating no significant global underprediction or overprediction. However, the slope parameter $\kappa$ for centrals is 0.623, suggesting a tendency toward underprediction at the high-SFR end and overprediction at the low-SFR end. Similar to the $M_\star$ case, most of these deviations occur in sparsely populated regions of parameter space. By contrast, the performance is substantially worse for satellites, with $\kappa = 0.138$. As is evident from the plot, the model struggles to reproduce the full dynamic range for satellites, instead collapsing toward a mean strip across the span. This is also reflected in the $R^2$ values: 0.692 for centrals, compared to only 0.19 for satellites.

An alternative strategy for predicting SFR used in this study is the fraction-based method. This approach estimates sSFR with MIG and then multiplies it by the MIG-predicted $M_{*}$ to derive the SFR. In the second layer of MIG, Q galaxies identified by the classifier are assigned a mean sSFR of $10^{-13.42},\mathrm{yr}^{-1}$. The predictions are shown in the second-row right panels of \autoref{fig:rmse_scatter}(a, b). For central galaxies, the $\sigma$ value from the sSFR method performs better than the direct SFR method, both in terms of accuracy ($R^2$ = 0.729) and bias in prediction ($\kappa$ = 0.734). For satellite galaxies, the sSFR-based method is highly effective, improving both accuracy ($R^2$ = 0.634) and bias ($\kappa$ = 0.605). On its own it may not seem very appealing, but this clearly shows how the fraction-based method can improve accuracy with the help of highly accurate $M_*$ predictions.

Thus, the fraction-based approach excels over the traditional feature-based method in predicting SFR, as evidenced by its higher accuracy and more uniform prediction distribution.

\subsection[HI mass]{$M_{\rm HI}$}
Predicting $M_{\rm HI}$ from halo features has received significant attention in ML applications to cosmology, due to the important role of $M_{\rm HI}$ in understanding galaxy formation and evolution, as well as in mapping the large-scale structure of the universe via 21cm emission-line surveys. Here, we outline the results for $M_{\rm HI}$ predictions from MIG.

As with SFR, predicting $M_{\rm HI}$ starts with a preclassifier to separate SF and Q galaxies, establishing a two-layer ML model framework for both central and satellite galaxies.

Subsequently, Q galaxies identified by the classifier are assigned a mean $M_{\rm HI}$ value of $10^{4.8}M_\odot$, while the $M_{\rm HI}$ for SF galaxies is predicted using the regressor layer of MIG. The results, shown in the third-row left panels of \autoref{fig:rmse_scatter}(a, b), yield a promising $R^2$ of 0.781 for central galaxies, but drop sharply to just 0.128 for satellites. In terms of distribution, the trend resembles the SFR predictions, with a better $\kappa$ of 0.689 for centrals, but a much weaker $\kappa$ of 0.141 for satellites.

We now turn to the fraction-based approach of MIG. This method involves predicting $M_{fHI}$, then multiplying it by the ML-predicted $M_{*}$ to derive $M_{\rm HI}$. Galaxies classified as Q were assigned a mean $M_{fHI}$ of $10^{-4.46}$, and the SF counterparts were estimated using the regressor layer of MIG. These $M_{fHI}$ predictions were then multiplied by the corresponding $M_{*}$ estimates, as discussed in \autoref{sec:sm}, to calculate $M_{\rm HI}$. The final predictions are shown in the third-row right panel of \autoref{fig:rmse_scatter}(a, b).

For central galaxies, $R^2$ shows an improvement over the traditional approach, at 0.792 and much better $\kappa$ of 0.799. For satellite galaxies, we again see a remarkable increase in the accuracy ($R^2$ = 0.453) and bias in prediction ($\kappa$ = 0.616). Visual inspection of \autoref{fig:rmse_scatter}(a) reveals a tighter spread for the $M_{fHI}$ method, particularly in the plot’s central region, where the dense yellow area appears slightly more centrally oriented toward the fraction-based approach, with a symmetric distribution across the best prediction line. \autoref{fig:rmse_scatter}(b) indicates that the fraction-based approach is much closer to an ideal prediction than the direct $M_{\rm HI}$ approach for satellite galaxies. So, we can establish a clear advantage for the fraction-based approach in accurately predicting $M_{\rm HI}$ values.

\subsection[H2 mass]{$M_{\rm H2}$}
Although predicting $M_{\rm H2}$ using ML has not attracted as much attention as $M_{\rm HI}$ predictions, it remains a crucial element for a comprehensive estimation of gas and baryonic content within galaxies. It will become increasingly important as CO line emission surveys for cosmology emerge \citep{keating2020intensity}. Below, we outline the pipelines developed for $M_{\rm H2}$ predictions within our ML framework and assess their accuracy.

In a manner similar to the SFR and $M_{\rm HI}$ predictions, estimating $M_{\rm H2}$ begins with a pre-classifier to separate SF from Q galaxies. After this step, Q galaxies are assigned a mean $M_{\rm H2}$ value of $10^{2.71}M_\odot$, while SF galaxies have their $M_{\rm H2}$ predicted using the regressor layer of MIG. Central galaxy predictions, shown in the fourth-row left panel of \autoref{fig:rmse_scatter}(a), demonstrate strong accuracy ($R^2$ = 0.806) and modest bias ($\kappa$ = 0.719). The high-end outliers also remain relatively close to the $y=x$ line. For satellite galaxies, however, the left panel of \autoref{fig:rmse_scatter}(b) reveals the same difficulties as before, with biased predictions that fail to cover the full distribution.

We further examine the fraction-based approach within our ML framework for predicting $M_{\rm H2}$. In this method, $M_{fH2}$ is first estimated using ML algorithms and then multiplied by the ML-predicted $M_{*}$ to recover $M_{\rm H2}$. Galaxies classified as Q are assigned a mean $M_{fH2}$ of $10^{-4.14}$, while SF galaxies are predicted using the regressor layer of MIG. The full results are presented in the fourth-row right panels of \autoref{fig:rmse_scatter}(a, b). For central galaxies, this approach yields a slight improvement in accuracy ($R^2$ = 0.813) but a much lower prediction bias ($\kappa$ = 0.821). The distribution is also notably symmetric about the $y=x$ line. For satellites, the fraction-based method again performs far better than the direct approach, with accuracy reaching $R^2$ = 0.804 and bias reduced to $\kappa$ = 0.72. The distribution is similarly symmetric and less biased, confirming that the fraction-based strategy clearly outperforms the direct method in predicting $M_{\rm H2}$.

\subsection[Metallicity Metallicity]{Metallicity $Z$}
Predicting metallicity is important for predicting galaxy photometry, as stellar emission depends on metallicity. It is also relevant for intensity mapping in metal emission lines such as [OII] and [OIII]. The metallicity of a galaxy is tightly linked to its stellar mass through the well-known mass–metallicity relation. Given this tight correlation, we anticipate that models predicting metallicity will achieve a high accuracy.

Following the methodology used for SFR, $M_{\rm HI}$, and $M_{\rm H2}$, the prediction of $Z$ also begins with a pre-classifier to separate SF from Q galaxies. This step provides the foundation for a two-layer ML framework, consistent across both central and satellite galaxies. Galaxies classified as Q are assigned a mean $Z$ value of $10^{-2}$, while SF galaxies have their $Z$ estimated using the regressor layer of MIG. Similar to $M_{*}$, the $Z$ predictions exhibit a very tight and strong correlation with the true values. The results, shown in the fifth row of \autoref{fig:rmse_scatter}(a, b), indicate an accuracy of $R^2$ = 0.665 and a low prediction bias ($\kappa$ = 0.708) for centrals. For satellites, the performance is even more impressive, with $\kappa$ = 0.905—the best among all predicted features—alongside a strong accuracy of $R^2$ = 0.752.

%%%%%%%%%%%%%%%%%%%%%%%%%%%%%%%%%%%%%%%%%%%%%%%%%%%%%%%%%%%%%%%%%%%%%%%%%%%%%%%%%%%%%%%%%%%%%%%%%%%%%%%%%%%%%%%%%%%%%%%%%%%%%%%%%%%%

\section[Predictions at high redshift]{Predictions at $z=1,2$}
\label{sec:regz}
Until now, our exploration of the MIG’s efficacy in predicting galaxy properties has been limited to $z=0$. However, to align our predictions with the broader scope of galaxy surveys or intensity mapping, which often extend to higher redshifts, it is crucial to adapt and evaluate our ML models at $z=1$ and $z=2$, using data from the {\sc Simba} 100 Mpc box.

\begin{figure}
  \centering
  \vspace{0.35cm}
  \includegraphics[width=\linewidth]{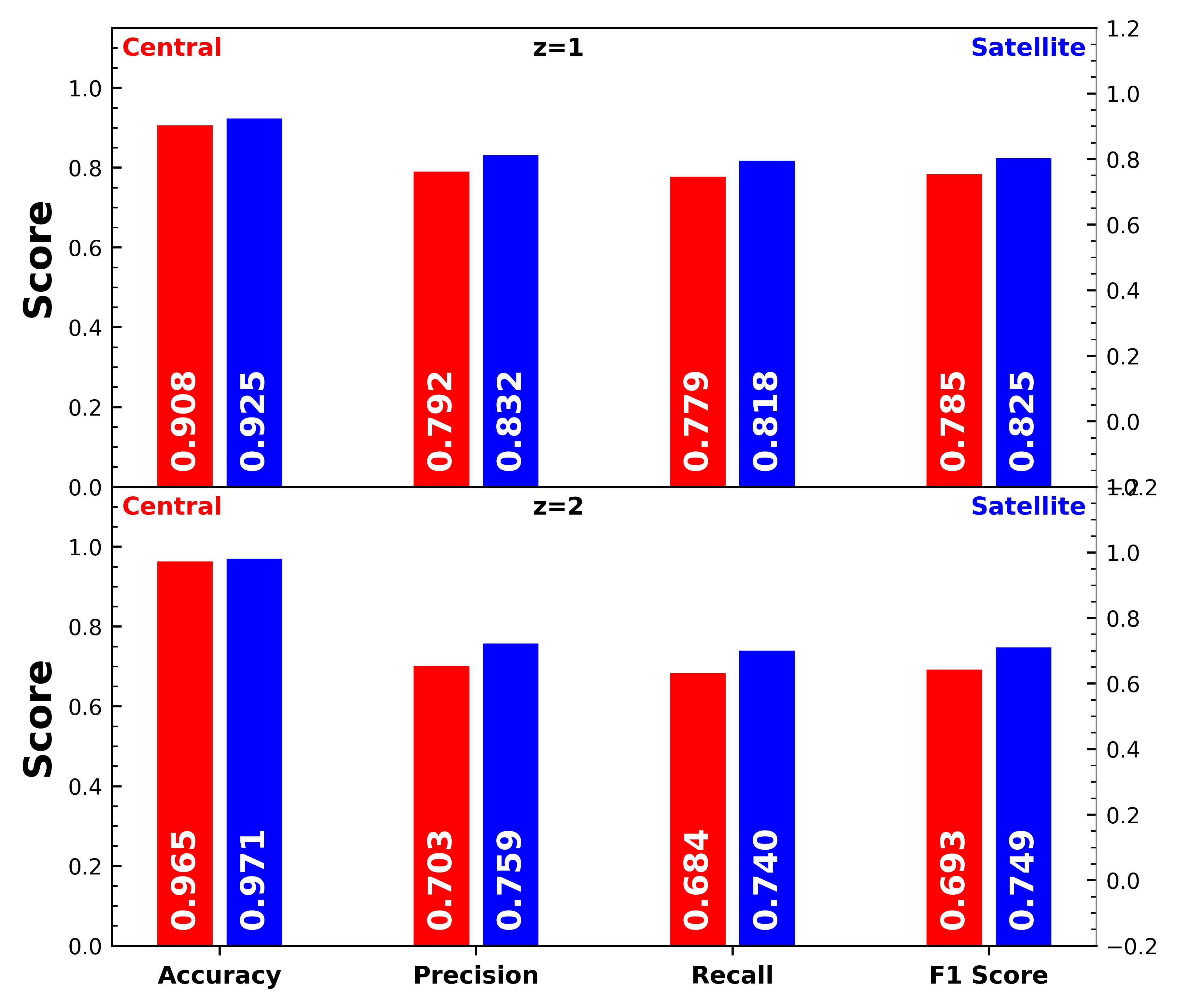}  
  \vspace{0.23cm}
  \caption{Performance metrics of the classifier layer of MIG for central (red) and satellite galaxies (blue) at $z=1,2$. The scores for each metric are indicated inside the respective bars.}
  \label{fig:clfz}
\end{figure}

In this context, we employ both fraction-based and feature-based approaches for predicting SFR, $M_{\rm HI}$, and $M_{\rm H2}$ at $z=1$ and $z=2$. As with the $z=0$ scenario, these predictions at higher redshifts begin with the application of a preclassifier, as illustrated in the classifier layer of \autoref{fig:flow}. This crucial step involves setting a boundary for sSFR, defined by \autoref{eq:sSFR}, which simplifies to $10^{-10.5}\,\mathrm{yr}^{-1}$ for $z=1$ and $10^{-10.2}\,\mathrm{yr}^{-1}$ for $z=2$. Following this guideline, the classifier layer of MIG is trained, and the metrics are shown in \autoref{fig:clfz}. MIG achieves high accuracy for both $z=1$ and $z=2$. Similarly to $z=0$, the central galaxies achieve higher accuracy compared to satellite galaxies across all metrics, although we notice a gradual decrease in the F1 score at higher redshifts. The simplest explanation for this is the decrease in sample size with increasing redshift. At $z=0$, we have 55,609 galaxies, whereas at $z=1$ and $z=2$ we have 39,298 and 30,743 galaxies, respectively. With the future addition of other simulations to MIG, we hope the accuracy will further improve at higher redshifts too.

\begin{figure*}
\begin{subfigure}{.45\textwidth}
  \centering
  \includegraphics[width=0.95\linewidth]{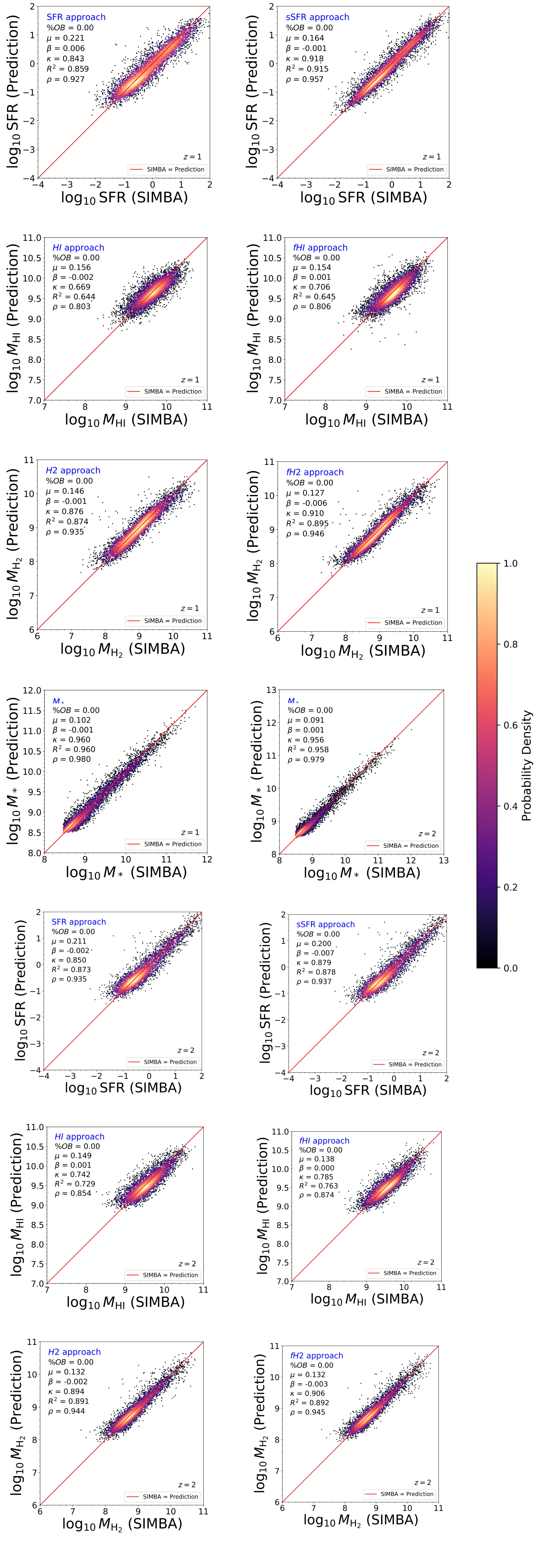}
  \caption{Central Galaxies}
\end{subfigure}
\begin{subfigure}{.45\textwidth}
  \centering
  \includegraphics[width=0.95\linewidth]{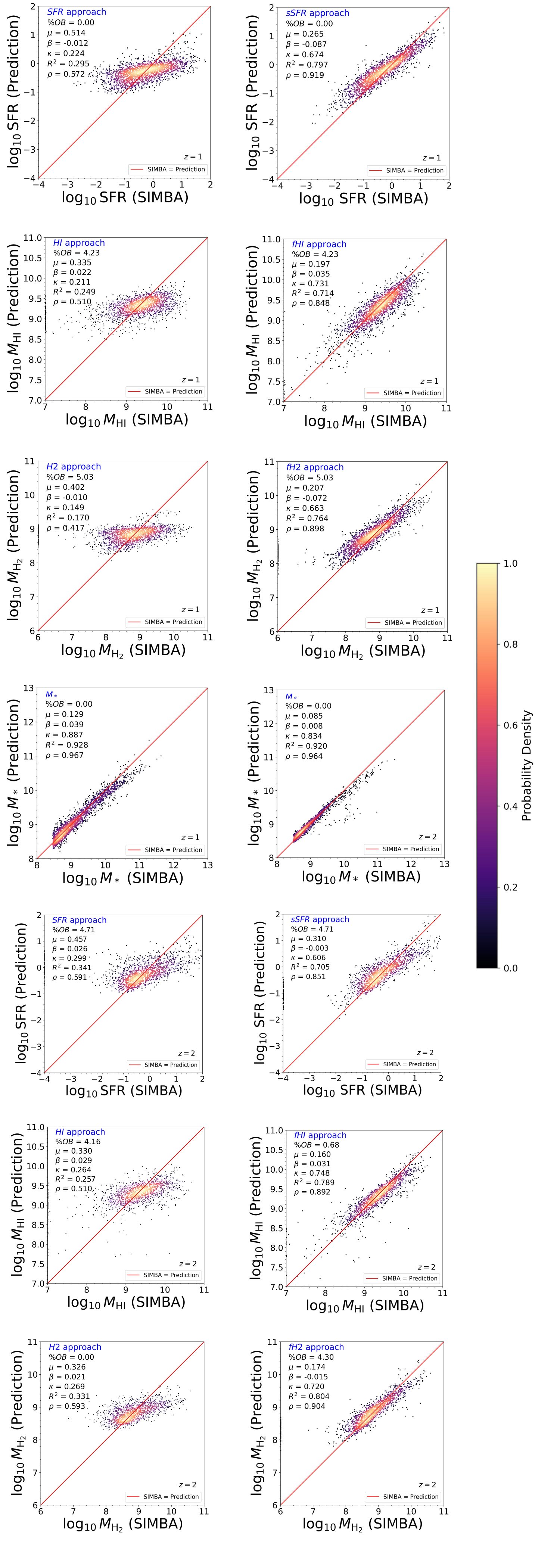}
  \caption{Satellite Galaxies}
\end{subfigure}
\caption{The best-performing frameworks for both fraction-based and feature-based approaches were implemented on the entire \simba simulation at $z=1$ and 2 to predict SFR, $M_{\rm HI}$, and $M_{\rm H2}$. (a) shows ML predictions for central galaxies, and (b) shows predictions for satellite galaxies. The top three rows are at $z=1$, and the bottom three rows at $z=2$ in both (a) and (b).}
\label{fig:scatterz}
\end{figure*}

We then implement the regressor layer of MIG in a manner similar to $z=0$. The predictions from these models are plotted against the true values in \autoref{fig:scatterz}(a, b), where we make a detailed comparison between the fraction-based and feature-only approaches for the prediction of the targeted features at $z=1$ (top three rows) and $z=2$ (bottom three rows). In the following subsections, we provide a detailed comparison of these approaches with respect to our model’s accuracy. We also present the mass functions for both approaches, similarly to $z=0$, in \autoref{fig:massfuncz}.

At $z=1$ and $z=2$, MIG shows a trend similar to that at $z=0$. For central galaxies, $M_{*}$ achieves strong accuracy with $R^2$ = 0.96 and 0.958, and only minimal bias in predictions ($\kappa$ = 0.96 and 0.956, respectively). Predictions for satellite galaxies are also strong, with $R^2$ = 0.928 at $z=1$ and 0.920 at $z=2$. At $z=2$, however, a few notable outliers appear, deviating significantly from the true values, though these occur only in low-density regions.

For SFR predictions, the fraction-based approach consistently outperforms the traditional method, delivering higher accuracy and lower bias for central galaxies at both redshifts. At $z=1$, the top panel of \autoref{fig:scatterz}(a) clearly shows that the fraction-based method is much better constrained, with a lower $\mu$ = 0.164 compared to 0.221 for the traditional approach. Satellites follow the same trend as at $z=0$, achieving much higher accuracy and reduced bias at both redshifts, primarily due to the highly accurate $M_{*}$ predictions. For $M_{\rm HI}$ predictions, we find a scenario similar to the SFR results for both central and satellite galaxies. For centrals, the predictions are more tightly constrained with reduced bias. For satellites, the traditional approach remains limited to a narrow central band and fails to capture the full trend, whereas the fraction-based method achieves much higher accuracy and significantly lower bias. For $M_{\rm H2}$ predictions, both approaches yield symmetric results with comparable accuracy for central galaxies at both redshifts. However, as with other properties, the fraction-based method performs better for satellites, successfully capturing the entire trend across both redshifts.

Overall, satellite galaxy predictions are consistently stronger with the fraction-based approach. For centrals, the difference is less pronounced, though the fraction-based method still maintains a clear advantage.

%%%%%%%%%%%%%%%%%%%%%%%%%%%%%%%%%%%%%%%%%%%%%%%%%%%%%%%%%%%%%%%%%%%%%%%%%%%%%%%%%%%%%%%%%%%%%%%%%%%%%%%%%%%%%%%%%%%%%%%
\section{Mass Functions}
\label{sec:massfunc}
\begin{figure}
    \includegraphics[width=\columnwidth]{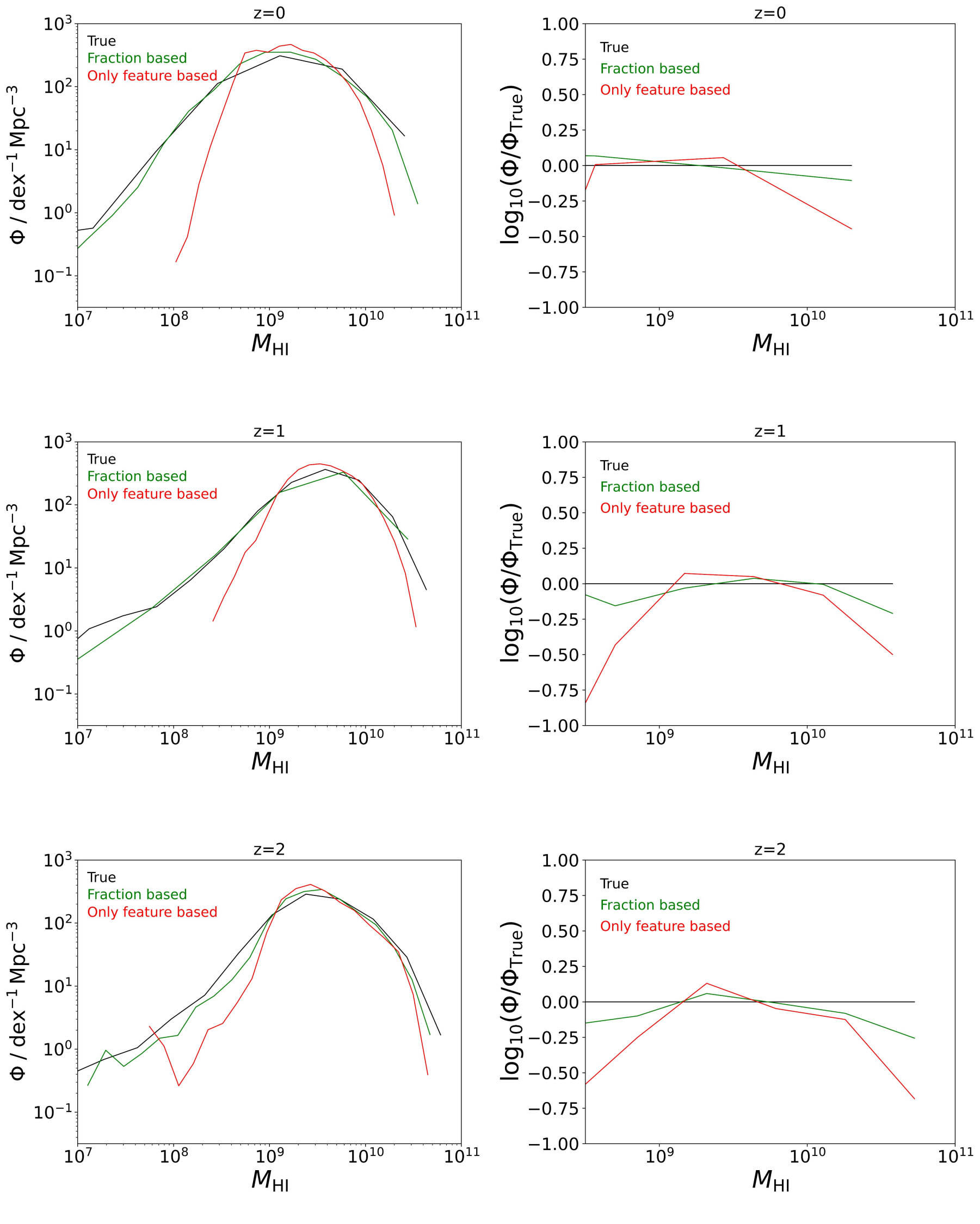}
    \caption{This figure shows mass functions of $M_{\rm HI}$ at $z$ = 0, 1, and 2. The left column has the true, fraction-based, and feature-based mass functions, while the right column shows the same values divided by the true values.}
    \label{fig:massfuncz}
\end{figure}

To accurately capture the baryonic mass distribution in galaxies, particularly within the framework of dark matter halo integration, our methodology employs several ML techniques as described in previous sections. However, as noted in \autoref{sec:Intro}, conventional ML models often suffer from under-dispersion, where the predicted distributions, measured through the mass functions, are narrower than the true distributions. In this section, we test whether both approaches in MIG can recover the true mass function as shown in \autoref{fig:massfuncz}. For this work, we focus on $M_{\rm HI}$ values from 100 Mpc/h volume at all the concerned redshifts, which is especially important for HI intensity mapping. Mass functions for other features can be presented upon request, but in general they follow a similar trend to $M_{\rm HI}$.

At all redshifts, as shown in the left panel of \autoref{fig:massfuncz}, the traditional approach produces a squeezed mass function, as expected. Since these mass functions are generated using all galaxies (both satellites and centrals), the effect of satellite predictions squeezing is clearly visible, with the model restricted to a narrow central strip. In contrast, the fraction-based approach yields a distribution much closer to the true mass function, though with slight underprediction at both the high and low mass ends and minor overprediction in the central region. This also matches our expectations, as most predictions tend to cluster around the central region. Nonetheless, the fraction-based approach captures the overall distribution far better than the traditional approach.

The main reason for highlighting Q galaxies in our work was to avoid compromising the regressor accuracy with arbitrary low values. Since Q galaxies contribute only a negligible fraction to the total $M_{\rm HI}$ in the simulation, their predictions are not of primary concern. The dominant contribution comes from galaxies with $M_{\rm HI} > 10^{7}M_\odot$, as shown in the right panel of \autoref{fig:massfuncz}. In this panel, we present the ratio of each approach relative to the true values to highlight biases in the predictions. We again observe that the traditional approach produces a more squeezed distribution, while the fraction-based method stands out as remarkably successful in reproducing the true mass function.

%%%%%%%%%%%%%%%%%%%%%%%%%%%%%%%%%%%%%%%%%%%%%%%%%%%%%%%%%%%%%%%%%%%%%%%%%%%%%%%%%%%%%%%%%%%%%%%%%%%%%%%%%%%%%%%%%%%%%%%%%%%%
\section{Feature importance}
\begin{figure*}
    \includegraphics[width=\textwidth]{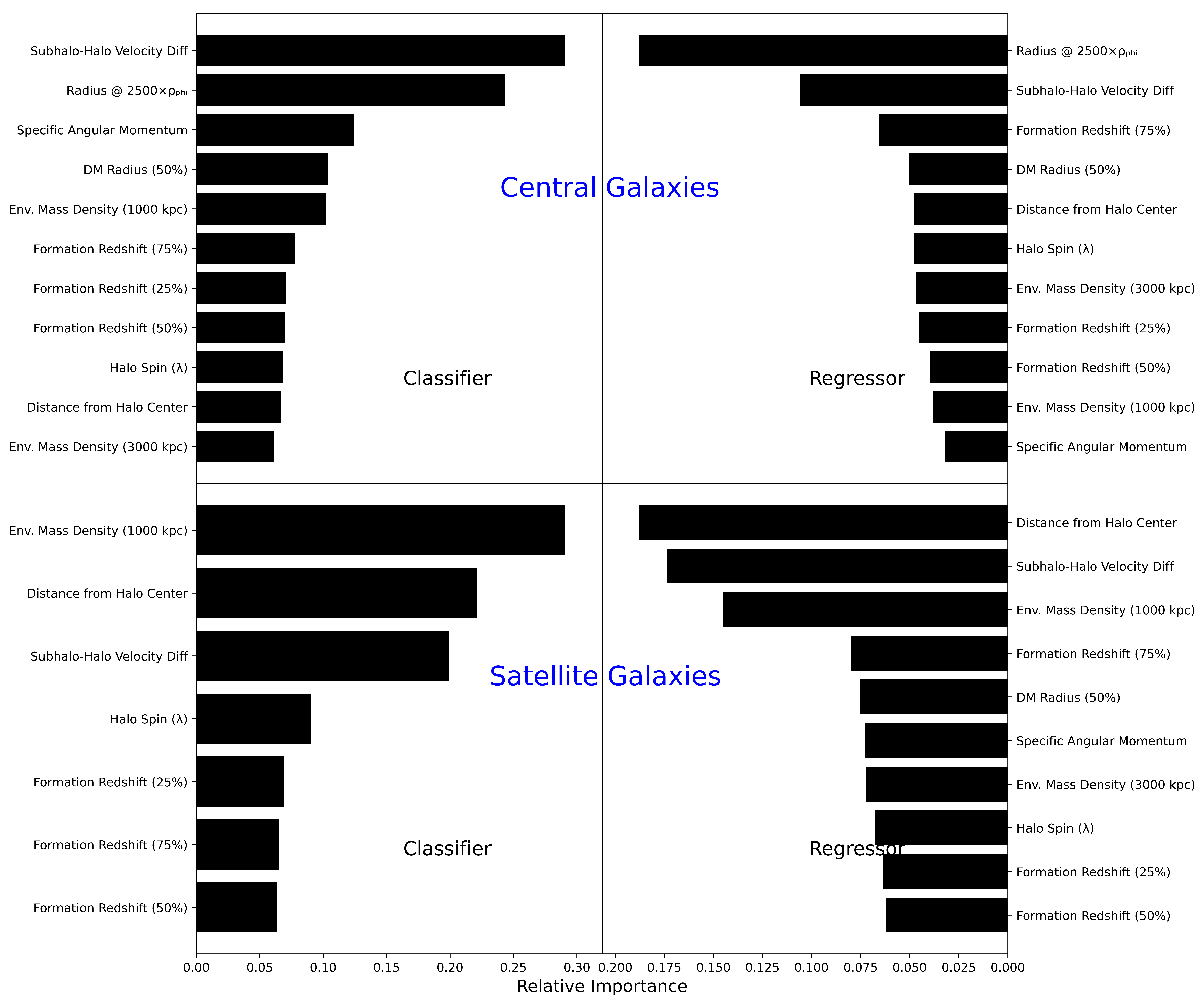}
    \caption{This figure shows feature importance as derived by the Random Forest (RF) algorithm. The top row has results for central galaxies, and the bottom row is for satellite galaxies. The left column represents a classifier’s feature importance, and the right column a regressor’s feature importance. The length of each bar corresponds to the relative importance of a specific halo property derived by the RF algorithm for inferring sSFR in galaxies using the feature-based approach.}
    \label{fig:feature_importance}
\end{figure*}

It is worthwhile to examine which input features drive the predictions to gain insight into the physics that MIG is learning from. As explained in \autoref{sec:properties}, we compute the correlation matrix using the publicly available Pandas library and identify features with a correlation coefficient exceeding 0.9. Among the correlated feature pairs, we remove the lower-ranked feature based on the Random Forest rankings, retaining only the most relevant and independent features. We do this first for the classifier layer, then for the regressor layer of each feature. Because MIG predicts quite a few galaxy properties, it is impractical to show all feature-importance plots in this paper. Hence, we discuss the feature importance for the classifier layer and the regressor layer for sSFR predictions.

The first step of feature selection for the MIG classifier at $z=0$ for central galaxies starts with building a correlation matrix among all the features. Here, we filter out features with an absolute correlation above 0.9. Then, we train a Random Forest classifier and get feature importance for each halo property. By keeping the high-importance feature from the correlated feature list, we end up with the top-left panel of \autoref{fig:feature_importance}.

To determine feature importance for our input features at $z=0$, we use the Random Forest algorithm available in the {\sc Scikit-Learn} library. Unfortunately, feature importance is only available using this ML method, so we cannot examine it for our entire pipeline; nonetheless, it may offer interesting insights. To provide a clear visual representation, \autoref{fig:feature_importance} selectively presents the feature importance only for the sSFR predictions.

Analyzing central galaxies and satellite galaxies shows that halo–subhalo connecting properties are the most significant features for the classifier model. Next in line are the radius measurements at various multiples of the density. These radii measurements tie into the galaxy’s spatial and mass distribution, affecting its gravitational potential and star formation rates.

%%%%%%%%%%%%%%%%%%%%%%%%%%%%%%%%%%%%%%%%%%%%%%%%%%%%%%%%%%%%%%%%%%%%%%%%%%%%%%%%%%

%%%%%%%%%%%%%%%%%%%%%%%%%%%%%%%%%%%%%%%%%%%%%%%%%%%%%%%%%%%%%%%%%%%%%%%%%%%%%%%%%%%%%%%%%%%%%%%%%%%%%%%%%%%%%%%%%%%%%%%%%%%%
\section{Comparison with other works}

A similar ML framework was introduced by \citet{10.1093/mnras/sty1169} using the \mufasa~\citep{10.1093/mnras/stw1862} simulation, but restricted to central SF galaxies. They achieved strong performance in predicting HI richness, with Pearson $r>0.8$ and RMSE $<0.3$ dex. In contrast, \citet{lovell2022machine} trained on EAGLE and C-EAGLE (including satellites) and reported that 82\% of galaxies had $M_*$ predictions within 0.2 dex, 79\% for SFR, and 65\% for gas mass. More recently, \citet{de2022mimicking} used IllustrisTNG300 to model both centrals and satellites, reaching $r \approx 0.98$ for $M_*$ but only $r \approx 0.7$–0.8 for SFR, sSFR, and size.

These comparisons show that while earlier works achieved high accuracy for stellar mass, Q and satellite populations remain challenging—often leading to compressed distributions or low $R^2$ in SFR and gas predictions. Our MIG framework addresses these gaps by explicitly separating SF and Q galaxies and including satellites, reaching $R^2$ values of $\sim$0.7–0.8 for $M_{\rm HI}$ and $M_{\rm H2}$ in centrals, and significantly improving satellite predictions with the fraction-based method. For Q galaxies, we take a pragmatic approach: assigning mean values rather than forcing the regressor into regions where the data are sparse. Since Q galaxies contribute negligibly to the global HI and H2 budget (dominating only below $10^{7}M_\odot$), this treatment avoids degrading the accuracy for the SF population, which drives the bulk of the signal relevant for HI intensity mapping. At the same time, our framework is flexible enough to provide estimates of all baryonic properties ($M_*$, SFR, $M_{\rm HI}$, $M_{\rm H2}$, $Z$, and fractions), making the Q treatment useful when full galaxy populations are needed.

%%%%%%%%%%%%%%%%%%%%%%%%%%%%%%%%%%%%%%%%%%%%%%%%%%%%%%%%%%%%%%%%%%%%%%%%%%%%%%%%%%%%%%%%%%%%%%%%%%%%%%%%%%%%%%
\section{SUMMARY}
In this work, our primary goal is to enable ML models to predict the baryonic properties of halos across an entire N-body simulation, independent of the galaxy’s location or type. Previous studies have often struggled to incorporate subgroups such as satellite and Q galaxies. Using the {\sc Caesar} catalogs, we first separate galaxies into centrals and satellites, and then train dedicated regressors for each subgroup to predict their baryonic content. To include Q galaxies in MIG, we apply a pre-classifier step and train the regressor only on the SF subset of the classified data, with Q galaxies assigned representative mean values. The final accuracy therefore reflects the combined performance of both ML layers. MIG achieves strong accuracy even for fluctuating properties that have traditionally been difficult to predict from halo properties alone, such as SFR, \HI\ mass, and H$2$ mass—particularly when using the fraction-based approach alongside accurate $M*$ predictions. At higher redshifts (\autoref{fig:scatterz}), the regressors perform better, reflecting the lower Q fraction in \simba\ at $z=1,2$ compared to $z=0$. However, this reduced Q population also makes the classifier more biased toward the SF set, leading to a modest decline in classification accuracy.

One interesting result is that in all cases our ML framework performed better when predicting the fraction relative to $M_*$ and then multiplying by the predicted $M_*$ (see \autoref{fig:rmse_scatter} and \autoref{fig:scatterz}). This may seem counterintuitive, since it requires predicting two separate quantities and combining them, but the ML models predict $M_*$ with such high accuracy that this does not reduce overall performance. In fact, working with fractions removes the simple ``bigger is bigger" scaling effect, allowing the models to focus on more subtle variations. For satellite galaxies, the fraction-based approach consistently outperforms the direct method across nearly all properties, particularly by reducing mass-dependent biases. At $z=1,2$, the advantage is less dramatic for centrals but still present, with results generally favoring the fraction-based approach. Overall, this makes the fraction-based method the preferred strategy for maximizing accuracy.

The main difference between central and satellite galaxies in our study is the ratio of SF to Q galaxies. At $z=0$, approximately 40\% of satellite galaxies are Q, in contrast to about 10\% for central galaxies. This difference significantly affects the accuracy of our classifier predictions by providing the satellite galaxies a less biased dataset to train on. 

Addressing this issue may involve developing more robust classifiers specifically for central galaxies by trying to give some weights to both subgroups, making the dataset unbiased. Our results derive from tests with several popular ML algorithms and ensembles, including both manual and automated methods for hyperparameter optimization using TPOT. While further improvements might be achieved by employing a more extensive ensemble of sophisticated ML algorithms, our current study establishes a baseline for future research. As the first analysis of both satellite and Q galaxies, we provide a benchmark for subsequent studies that aim to use fully automated ML models across entire simulations to generate intensity maps, rather than focusing solely on subgroups.

A critical aspect of our methodology is the recovery of the mass function, a key metric for intensity mapping. Conventional ML models often learn best from the high-density mid-range but treat the low and high mass ends as outliers, leading to an artificially narrowed mass function. In contrast, our fraction-based approach closely reproduces the true mass function and performs significantly better than the traditional method (\autoref{fig:massfuncz}).

Beyond these broad results, some detailed key findings can be summarized as follows:
\begin{itemize}
    \item Among the input features selected from SIMBA, halo–subhalo properties consistently emerged as the most important for both centrals and satellites. For centrals, velocity offsets and halo radii were dominant, while for satellites, environmental density and proximity to the host halo played the strongest roles. Additional features such as halo spin, formation redshift, and dark matter radius also contributed, but with comparatively lower importance.
    \item The fraction-based approach significantly improved results for satellite galaxies and showed considerable improvement for central galaxies compared to the traditional approach across redshifts.
    \item Mass functions clearly highlight the advantages of the fraction-based approach. Across nearly all cases, this method consistently recovers the full spread of the true mass function, whereas the traditional approach tends to produce a narrower distribution.
\end{itemize}

A principal application of our ML framework is to simulate \HI intensity maps. Our frameworks can generate \HI intensity maps tailored for upcoming astronomical surveys such as the Hydrogen Intensity and Real-time Analysis eXperiment (HIRAX)~\citep{newburgh2016hirax} and the Square Kilometre Array (SKA)~\citep{bacon2020cosmology}. This can be done by applying our framework to large-scale N-body simulations with sufficient volume to obtain a cosmologically representative sample of large-scale structure, yet with enough resolution to properly resolve the lower-mass halos dominating the \HI\ content.

We further aim to improve the ML predictions for other galaxy quantities, such as emission-line surveys that track the star formation rate and CO intensity mapping surveys that trace molecular hydrogen mass. More advanced methods are needed for the classifiers to boost overall accuracy, progressing toward fully automated ML models for entire simulations. With a framework that can handle both centrals and satellites and reliably distinguish SF gas-rich populations, numerous applications arise for making predictions for future cosmology surveys, thus constraining both dark energy and the physical processes of galaxy evolution.

%%%%%%%%%%%%%%%%%%%%%%%%%%%%%%%%%%%%%%%%%%%%%%%%%%%%%%%%%%%%%%%%%%%%%%%%%%%%%%%%%%%%%%%%%%%%%%%%%%%%%%%%%%%%%%%%%%%%%%%%%%%
\section*{Acknowledgements}
PKD acknowledges the help of Dr. Tuhin Ghosh in accessing the Aquila cluster at NISER, Bhubaneswar, supported by the Department of Atomic Energy of the Government of India, to run various codes of the project. PKD also thanks Mr. Kali Charan Baksi for useful ML discussions. RD acknowledges support from STFC AGP Grant ST/Y001117/1. RD also thanks the Stellenbosch Institute for Advanced Studies Fellows program. WC is supported by the STFC AGP Grant ST/V000594/1, the Atracci\'{o}n de Talento Contract no. 2020-T1/TIC-19882 granted by the Comunidad de Madrid in Spain, and science research grants from the China Manned Space Project. WC also thanks the Ministerio de Ciencia e Innovación (Spain) for financial support under Project grant PID2021-122603NB-C21 and the HORIZON EUROPE Marie Sklodowska-Curie Actions for supporting the LACEGAL-III project with grant number 101086388.

%%%%%%%%%%%%%%%%%%%%%%%%%%%%%%%%%%%%%%%%%%%%%%%%%%%%%%%%%%%%%%%%%%%%%%%%%%%%%%%%%%%%%%%%%%%%%%%%%%%%%%%%%%%%%%
\section*{Data Availability}
The article’s data will be shared upon reasonable request to the corresponding author. The \simba\ simulation~\citep{dave2019simba} data used in this article is available at \url{http://simba.roe.ac.uk/}.


\bibliographystyle{mnras}
\bibliography{simbaML} % if your bibtex file is called example.bib

%%%%%%%%%%%%%%%%%%%%%%%%%%%%%%%%%%%%%%%%%%%%%%%%%%
%%%%%%%%%%%%%%%%% APPENDICES %%%%%%%%%%%%%%%%%%%%%

\appendix

\section{Generating Scatter plots}
\label{sec:app1}
\begin{figure}
\begin{subfigure}{.24\textwidth}
  \centering
  \includegraphics[width=1\linewidth]{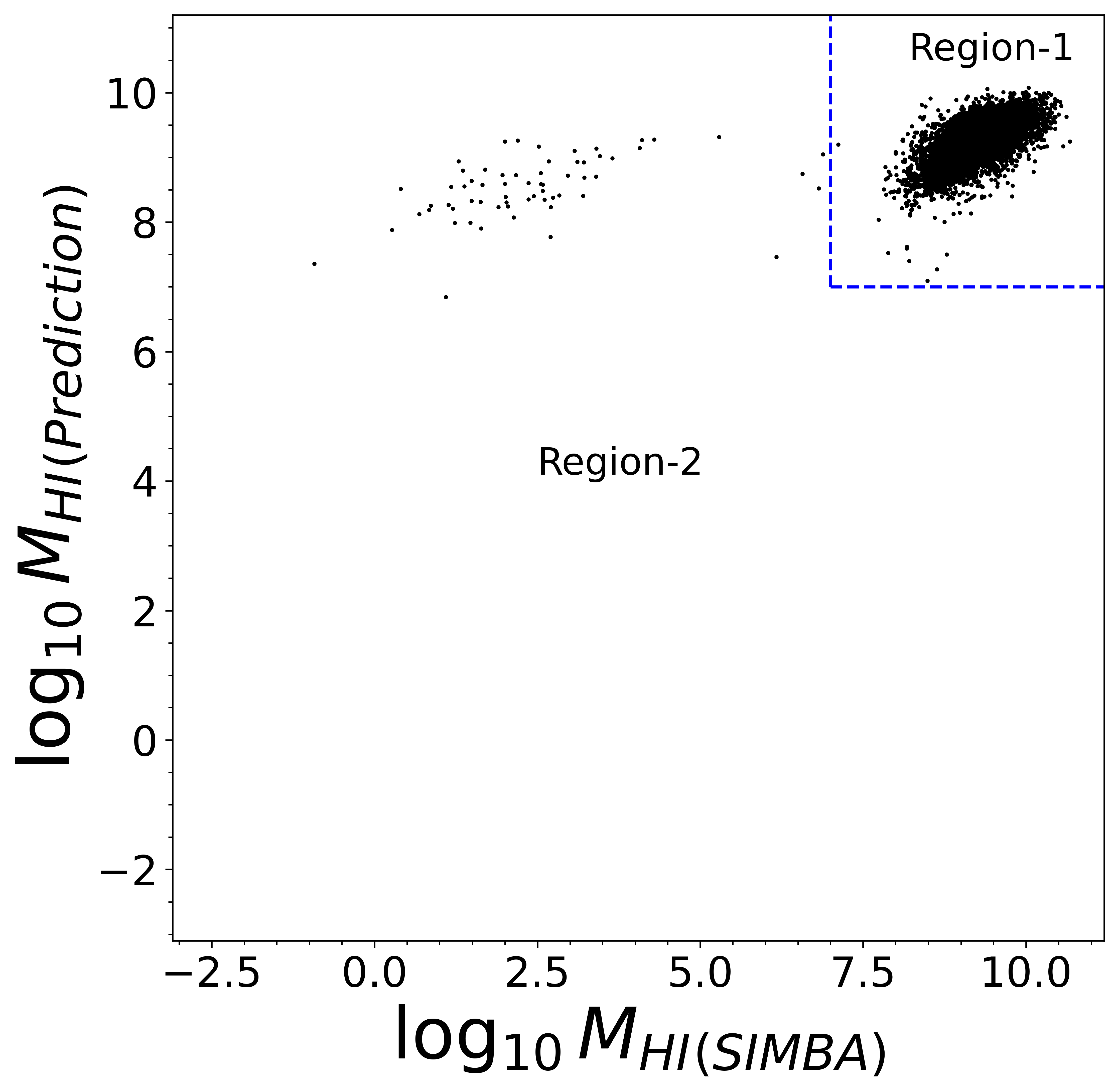}
  \caption{}
\end{subfigure}
\begin{subfigure}{.24\textwidth}
  \centering
  \includegraphics[width=1\linewidth]{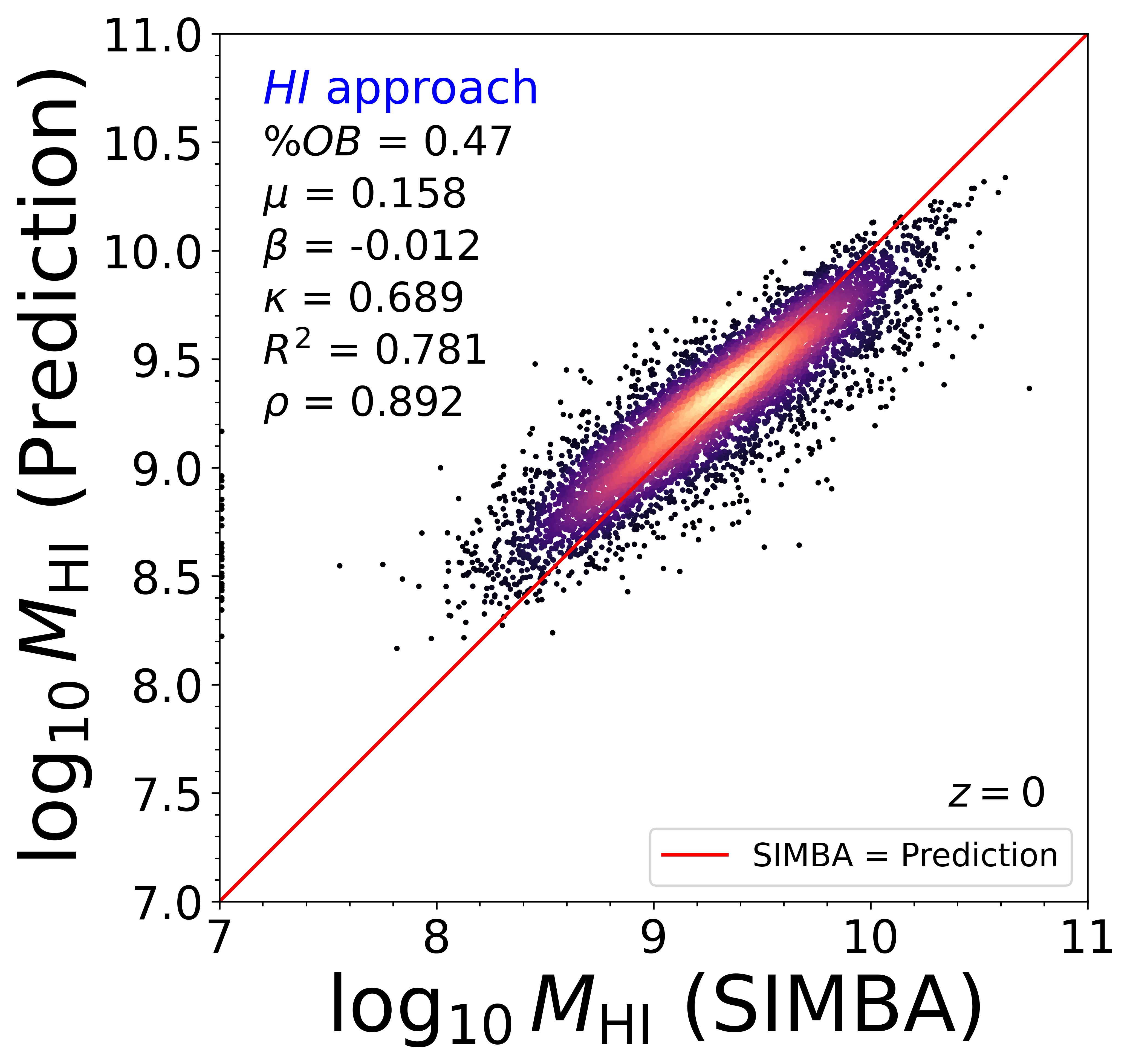}
  \caption{}
\end{subfigure}
\caption{Here we show the process of making the scatter plots from the predicted data in our paper. We have taken the example of the $M_{\rm HI}$ approach at $z=0$. (a) shows the final ML predictions using the $M_{\rm HI}$ approach at $z=0$ on the test data, plotted against the true (\simba) values. The entire graph is divided into two regions (Region-1 and Region-2) by dotted blue lines, drawn at the Q–SF boundaries. Region-1 has the SF data points from \simba\ that were predicted as SF. Region-2 has SF data points from \simba\ that were predicted as Q. (b) is the final plot used in the paper, which has the data points in Region-1; it also includes data points from neighboring regions, shifted to the nearest point on its boundary. The plots also include additional information in the corners along with a density-plot representation, to visualize the distribution more clearly.}
\label{fig:append}
\end{figure}

\autoref{fig:append} illustrates the process of creating these scatter plots, using $M_{\rm HI}$ as an example, although the same method applies to the other properties. The full distribution of all galaxies is shown in panel (a). This is divided into four regions. Region 1 is where both \simba\ and the ML indicate substantial \HI. Region 2 shows galaxies that are Q in \simba\ but predicted as SF by the ML, basically the extreme outliers from MIG.

From the entire \simba\ simulation at $z=0$, Region 1 of panel (a) has around 99.9\% of the global $M_{\rm HI}$. Thus, these misclassifications do not significantly affect the overall \HI budget. This explains why we will primarily be concerned with Region 1, which is shown alone in panel (b), with a 1:1 line shown in red. Since the points become saturated, we convert to a heat map of probability density, shown in panel (c). We further show all computed metrics in the upper left. In the following sections, we assess ML performance using figures like panel (c) for all quantities and redshifts.

% Don't change these lines
\bsp    % typesetting comment
\label{lastpage}
\end{document}